\documentclass[aps,prx,reprint,superscriptaddress,twocolumn,notitlepage,floatfix,longbibliography]{revtex4-2}
\usepackage{tikz,enumitem}
\usetikzlibrary{calc,arrows,decorations.pathreplacing,backgrounds}
\colorlet{mylinkcolor}{blue!66!black!80}
\usepackage[colorlinks=true,linkcolor=mylinkcolor,citecolor=mylinkcolor,filecolor=cyan,urlcolor=mylinkcolor,breaklinks=true]{hyperref}
\usepackage{xcolor}

\usepackage[utf8]{inputenc}
\usepackage[english]{babel}
\usepackage{amsmath}
\usepackage{amssymb}
\usepackage{bm,bbm}
\usepackage{hyperref}
\usepackage{xcolor}
\usepackage{graphicx}
\usepackage{tikz}
\usepackage{type1cm}
\usepackage{float}
\usepackage{amsthm}
\newcommand{\angstrom}{\mbox{\normalfont\AA}}

\begin{document}
\def\bR{\mathbf{R}}
\def\bC{\mathbf{c}}
\def\bS{\mathbf{s}}
\def\bB{\mathbf{B}}
\newcommand{\M}{{\textsf M}}
\newcommand{\Hess}{{\textsf H}}
\newcommand{\HS}{{\textsf S}}
\newcommand{\HB}{{\textsf B}}
\newcommand{\HC}{{\textsf C}}
\newcommand{\HM}{{\textsf M}}
\newcommand{\HY}{{\textsf Y}}
\newcommand{\Po}{\hat{\rm P}}
\def\T#1{#1^\mathrm{T}}
\def\bet{\boldsymbol{\eta}}
\def\bq{\boldsymbol{\zeta}}
\def\btau{\boldsymbol{\tau}}
\def\brho{\boldsymbol{\rho}}
\def\be{\mathbf{e}}
\def\bq{\mathbf{q}}
\def\br{\mathbf{r}}
\def\bc{\mathbf{c}}
\def\bs{\mathbf{s}}
\def\bx{\mathbf{x}}
\def\by{\mathbf{y}}
\def\bv{\mathbf{v}}
\def\m#1{\mathrm{#1}}
\def\T#1{#1^\mathrm{T}}
\def\du#1{\mathrm{\underline{\underline{#1}}}}
\def\Trans#1{#1^\mathrm{T}}
\def\kb{k_{\m{B}}}
\def\peq{P_{\m{E}}(\bq)}

\newtheorem*{Definition}{Hypothesis}

\definecolor{forestgreen}{HTML}{1A771A}
\def\note#1{{\small{\textcolor{forestgreen}{(#1)}}}}
\def\todo#1{{\color{red}(todo: #1)}}

\newcommand{\TODO}[1]{\textcolor{red}{TODO: #1}}
\newcommand{\DONE}[1]{\textcolor{forestgreen}{DONE: #1}}
\newcommand{\blue}[1]{\textcolor{black}{#1}}
\newcommand{\red}[1]{\textcolor{black}{#1}}

\author{Maximilian Vossel}
\affiliation{Mathematical bioPhysics group, Max Planck Institute for Multidisciplinary Sciences, G\"{o}ttingen 37077, Germany}
\author{Bert L.~de Groot}
\affiliation{Computational Biomolecular Dynamics group, Max Planck Institute for Multidisciplinary Sciences, G\"{o}ttingen 37077, Germany}
\author{Alja\v{z} Godec}%
\email{agodec@mpinat.mpg.de}
\affiliation{Mathematical bioPhysics group, Max Planck Institute for Multidisciplinary Sciences, G\"{o}ttingen 37077, Germany}%



\title{
The allosteric lever: towards a principle of specific allosteric response}

\date{\today}
\begin{abstract}

  Allostery, the phenomenon by which the perturbation of a molecule at one
  site alters its behavior at a remote functional site,
  enables control over biomolecular function.~Allosteric modulation
is a promising 
avenue for drug
discovery and 
is employed in the design of mechanical metamaterials.~However,
\blue{a general principle
of allostery,~i.e.\ a set of
quantitative and transferable ``ground rules'',}
remains elusive.~It is neither a set of structural
motifs nor intrinsic motions.~Focusing on elastic network models, we
here show that an \emph{allosteric lever}---a mode-coupling pattern induced by the perturbation---governs the
directional, source-to-target, allosteric communication:~a structural perturbation of an allosteric 
site
couples the excitation of localized
hard elastic modes with concerted long range
soft-mode relaxation.~Perturbations of non-allosteric sites instead
couple hard and soft modes uniformly.~The allosteric response is
shown to be generally non-linear and non-reciprocal, and
allows for minimal structural distortions
to be
efficiently transmitted to specific changes at distant
sites.~Allosteric
levers exist in proteins and ``pseudoproteins''---networks designed to display an
allosteric response.~Interestingly, protein sequences
that constitute allosteric transmission channels are shown to be evolutionarily
conserved. To illustrate how the results may be applied in
drug design, we use them to 
successfully predict known allosteric sites in
proteins.
\end{abstract}  
\maketitle

Allosteric proteins display specific responses of functionally
active sites to the binding of ligands at, typically remote, allosteric
sites \cite{Changeux_Edelstein_Allosteric_2005}.
With the exception of dynamic, entropy driven 
allosteric systems~\cite{Tsai_delSol_Nussinov_Allostery_2008,
McLeish_Schaefer_Heydt_allosteron_2018,Cooper_Dryden_Allostery_1984,Ozkan}, the 
allosteric response involves conformational
changes~\cite{Mccammon_Gelin_Karplus_Wolynes_hinge_1976,Daily_Gray_Local_2007,Zheng_Brooks_Thirumalai_Low_2006}. These
have been studied
experimentally~\cite{Laskowski_Gerick_Thornton_Structural_2009,NMR,Strain,Hamm}
(meanwhile also on the level of individual molecules \cite{Haran,Haran_1,Haran_2,Haran_3,Haran_4,Hagen_2019,Hagen_DNA}), using computational
approaches~\cite{Laine,Sali,Smith_Groot_Allosteric_2016,Graeter_oneway,Stock_1,Delemotte,Graeter23,Straub}
(also combined with experiments \cite{Stock}), as well as
theoretically by means of graph
theory~\cite{Amor_Schaub_Yaliraki_Barahona_Prediction_2016,
delSol_Fujihashi_Amoros_Nussinov_Residues_2006,Kaya} or distance
geometry~\cite{Greener_Filippis_Sternberg_Predicting_2017}, and statistical structure analysis \cite{Atilgan_screened,Mitternacht}. 
Allostery is not limited to proteins; DNA \cite{Garcia,Kim_13,Hagen_2019,Hagen_DNA}, RNA
\cite{Peselis_15,Ozkan_2}, and complexes of proteins and RNA \cite{Walker_20}
display allostery as well.
Allosteric control over
conformations is a promising new avenue for drug
discovery
\cite{Nussinov_Tsai_Allostery_2013,Chatzigoulas_Cournia_Rational_2021}
and is furthermore employed to design mechanical metamaterials with
complex responses
\cite{Nagel,Yan_Ravasio_Brito_Wyart_Architecture_2017,Rocks_Pashine_Bischofberger_Goodrich_Liu_Nagel_Design_2017,Kim_Lu_Strogatz_Bassett_Conformation_2019,PRX_new}.  
Allosteric concepts were recently extended to the control of flow
in complex networks \cite{Flow}. 

Since the pioneering discovery of switch-like behavior in oxygen
binding to hemoglobin \cite{Bohr}, our
understandind of allostery as a general \emph{action at a distance}
phenomenon came a long way. From the early phenomenological
``population shift'' 
\cite{Monod} and 
``induced fit'' models 
\cite{Koshland}, as well as Eigen's generalization combining both
\cite{Eigen}, which all relied heavily on static crystallographic structures of
proteins, the focus gradually shifted towards understanding the
microscopic, dynamic basis of the effect. Moreover, whereas initially
only multimeric proteins were thought to display allostery, 
it was later also found in monomeric proteins \cite{Ascenzi}. 
Despite successful applications of these early models and their
various recent generalizations and refinements \cite{landscape,Hilser,Motlagh}, phenomenological models have 
shortcomings. In particular, the structural coupling between the
allosteric source and functional sites 
\emph{cannot} be
resolved with a purely thermodynamic approach
\cite{Cui,Kar,Thirumalai,Hagen_Rev}.

Notwithstanding individual successes in explaining the structural determinants of
allostery in well-documented systems \cite{Daily_Gray_Local_2007,Nussinov_Tsai_Allostery_2013,Wodak_etal_Allostery_2019} and in designing
mechanical metamaterials
\cite{Yan_Ravasio_Brito_Wyart_Architecture_2017,Rocks_Pashine_Bischofberger_Goodrich_Liu_Nagel_Design_2017,Flechsig_Design_2017,Kim_Lu_Strogatz_Bassett_Conformation_2019,PRX_new},
a general mechanism that gives rise to allostery still eludes
understanding \cite{Fenton,Hilser,Motlagh,Hagen_Rev}.  
Functional allosteric motions are specific
\cite{Daily_Gray_Local_2007}, generally non-linear \cite{Miyashita_Onuchic_Wolynes_Nonlinear_2003,nonlinear_1,nonlinear_2}, 
and require concerted rearrangements
\cite{Daily_Gray_Local_2007}.~In contrast to simply cooperative networks
displaying single soft-mode responses \cite{Yan_Ravasio_Brito_Wyart_Principles_2018},
functional responses of proteins typically require coordinated motions that depend on the
network structure in a complex
manner~\cite{Miyashita_Onuchic_Wolynes_Nonlinear_2003,Kim_Lu_Strogatz_Bassett_Conformation_2019}. As
a result, it is much easier to predict the response to a given
perturbation \cite{Flechsig_Design_2017,nonlinear_1,nonlinear_2,Poma_Li_Theodorakis_Generalization_2018} than to identify the source site that upon
perturbation yields a specific allosteric response.   
A general strategy
for identifying allosteric source-target pairs that carry a desired, functional response
remains elusive.

\begin{figure*}[ht]
  \includegraphics[width=1.\textwidth]{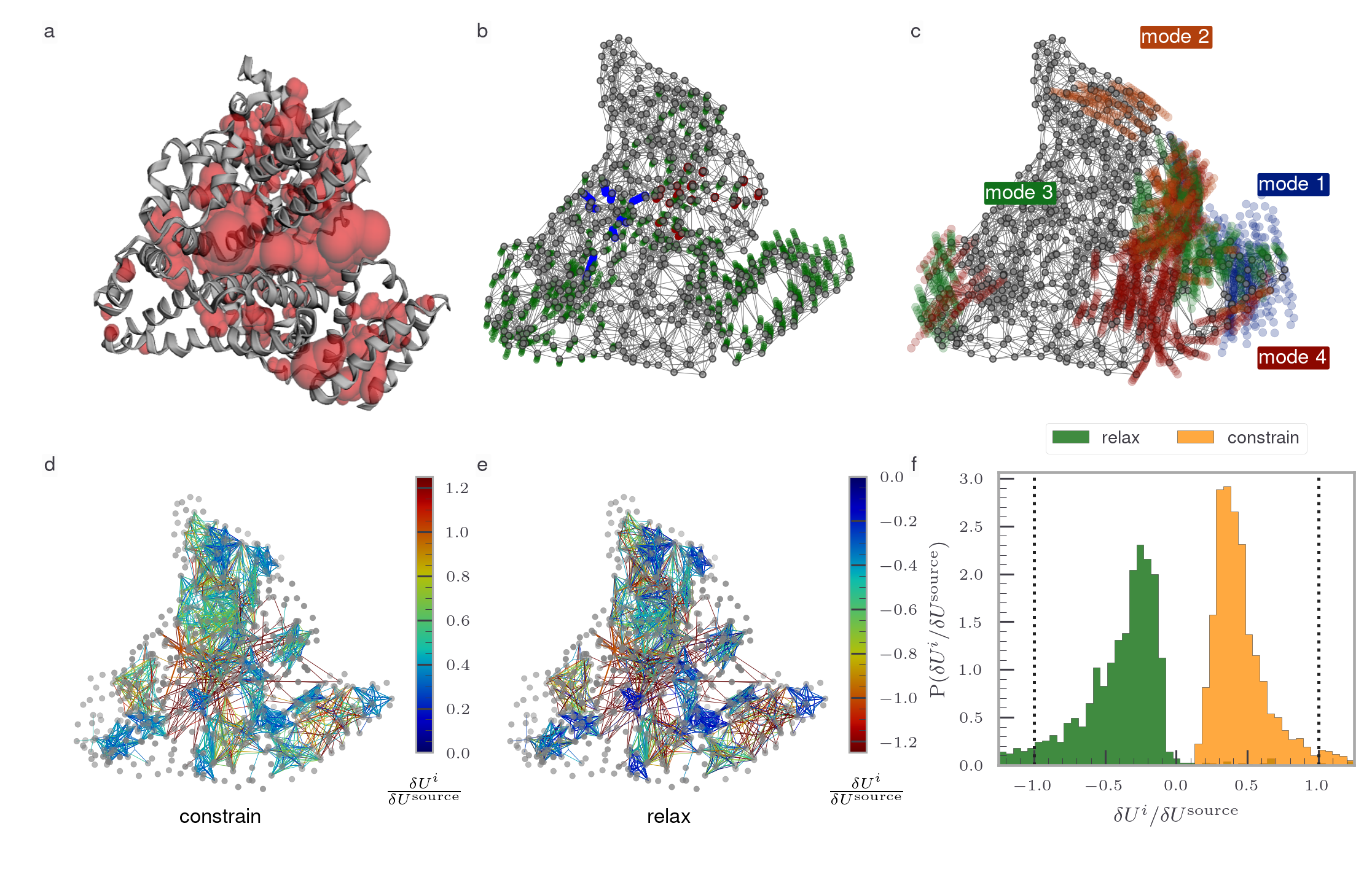}%
  \caption{
\textbf{Mechanical perturbation of binding-pocket candidates reveals unique
  character of allosteric sites.} (\textbf{a}) 
Cartoon representation of the Human Serum Albumin protein (HSA, PDB ID:
2bxd~\cite{ghuman2005structural}) showing binding-pocket candidates
(red) determined based on geometrical arguments
\cite{tian2018castp}. Only the largest 9 of the total of 77
analyzed pockets are shown.~(\textbf{b})~Elastic network model of the protein in \textbf{a}, coarse
grained on a residue level;~the beads are centered at carbon-$\alpha$
positions. The 
input perturbation at the allosteric source site and the
corresponding nonlinear response are shown in blue and green,
respectively; The response of beads in the biologically active pocket
(i.e.\ ``target'') is highlighted in red. The source and
target site are the known binding sites of warfarin and heme, respectively.
(A movie of the response trajectory is given in the SM.) (\textbf{c})
The four softest nonzero eigenmodes of the Hessian of the unconstrained
network in (\textbf{b}) superimposed on the structure 
(only substantial motions are shown); 
Note the complete lack of soft-mode participation in the response of
source-pocket beads.~(\textbf{d}-\textbf{f})
Energy changes accompanying contractions of binding-pocket candidates
during loading (\textbf{d}) and relaxation (\textbf{e}) 
The colors denote the energy change $\delta U^i$ upon contraction of the $i$th pocket 
(\textbf{d}) and during the corresponding relaxation (\textbf{e})
relative to that of closing the actual
source binding site $\delta U^{\rm source}$.~(\textbf{f})~Statistics
over all source-pocket candidates for the \blue{constraining}
and relaxation steps presented as green are orange histograms,
respectively. Dotted lines indicate  $\delta U^{\rm source}$ of
closing the true source pocket.  
Note that only a few pockets show substantial energy
changes. Allosteric source sites seemingly enable the largest energy uptake
and transmission upon mechanical perturbations.}
\label{fig:example_intro}
\end{figure*}

Minimal Elastic Network Models~\cite{Tirion_Large_1996}
have proven invaluable for studying allostery in numerous 
proteins~\cite{Zheng_Brooks_Identification_2005,
  Zheng_Brooks_Thirumalai_Low_2006} and for designing mechanical
networks with programmed responses \cite{Yan_Ravasio_Brito_Wyart_Architecture_2017,
Rocks_Pashine_Bischofberger_Goodrich_Liu_Nagel_Design_2017,Flechsig_Design_2017}.~However,
even here
a general understanding of the transmission of an allosteric signal from a 
source site into a specific, functionally relevant rearrangement of
the target site \cite{Daily_Gray_Local_2007} so far was 
out of reach.

From a mechanical perspective it is well understood how the curvature of the potential energy
surface
encodes collective motions---the eigenmodes of the Hessian---of elastic networks, which
was fruitfully exploited in the design of ``simply cooperative'' networks
\cite{Yan_Ravasio_Brito_Wyart_Architecture_2017,Ravasio_Flatt_Yan_Zamuner_Brito_Wyart_Mechanics_2019}. Nevertheless, 
a specific rearrangement of the target site must generally involve a balanced
interplay of many collective modes. Concurrently,
a given input perturbation of an allosteric source site
should intuitively ``dissipate'' the minimal amount of the perturbation energy near the source
in order to allow substantial long-distance rearrangements at the
target site.

The difficulty in elucidating generic allosteric mechanisms is rooted
in the fact that
although some allosteric responses are classifiable into intuitively understandable motions,
e.g.\ hinge, shear, piston-like, twist, rocking or combinations of
these, most proteins resist such a classification
\cite{Gerstein,Taylor}. This is due to the
nonexistence of discrete categories of motions; proteins can choose from a
continuum of different mechanisms
\cite{Liu_Doing_2021}.
It is therefore conceivable that
a generic ``physical principle of allostery'', if existent, is not a
set of common motion patterns. Instead it 
rather seems to be
\emph{a particular ligand-binding induced coupling pattern   
of many
complex 
modes} 
  that enables
an efficient channeled transmission of elastic
perturbations in a network.

By analyzing 14 different allosteric
proteins with known source-target pairs deposited in the
Protein Data Bank and generating a set of 30 artificial
``pseudoproteins'' trained
to display a specific allosteric response, we here provide 
evidence that fully supports the above hypothesis.
\blue{While previous works
	\cite{hawkins2006coupling}
have already explored allosteric communication and coupling patterns of normal
modes using the ENM approach, our hypothesis introduces a unique and
distinct coupling pattern where the source pocket excites hard modes
and the target pocket relaxes along soft modes. This is in contrast
to previous works, where the coupling pattern goes from soft to
stiff modes. In addition, our work focuses primarily on the mechanical
aspects of this coupling pattern, in contrast to earlier studies that
emphasized thermodynamic consequences. Thereby it offers a
complementary perspective to existing research in the field.}

We here propose a
conceptual shift, the \emph{allosteric lever}---a structural perturbation of
the allosteric source site couples hard and soft anharmonic elastic
modes, which allows for an efficient, directed 
(i.e.\ non-reciprocal), and specific transmission to a distant target site. 
An extensive analysis of the energetics and local
spectral properties of the response confirms our idea. Moreover,
protein sequence patterns involved in allosteric transmission channels are found to be evolutionarily
conserved. We use the allosteric-lever principle to 
successfully predict known allosteric sites in
proteins.

The manuscript is structured as follows.
In Sec.~\ref{setup} we describe the model and its parameterization and
present a new method to determine the full (i.e.\ beyond linear) allosteric response. In Sec.~\ref{Lever} we analyze the
energetics of allosteric responses and formulate the \emph{allosteric
lever} hypothesis. Next, in Sec.~\ref{nonreci} we establish that
allosteric responses are in general  non-linear and non-reciprocal. In
Sec.~\ref{sec_specific} we show that true (functional) allosteric responses
are specific, and in Sec.~\ref{sec_predictive} we demonstrate how already the most
basic version of our findings may be utilized to predict allosteric 
binding sites in proteins. In Sec.~\ref{evo} we find that protein
sequence patterns that comprise allosteric transmission channels are
evolutionary conserved. 
We conclude with a perspective and give an outlook on future
research directions that are provoked by our findings. 




\section{Setup and mechanical response}\label{setup}

\subsection{Model and parameterization}

Throughout we focus on the Elastic Network Model (ENM) used to describe the
long-timescale dynamics of proteins on the coarse-grained level of amino
acids \cite{Haliloglu_Bahar_Erman_Gaussian_1997}. The nodes of the
network are beads connected by Hookean springs with stiffness $\kappa$ (see
Fig.~\ref{fig:example_intro}). Denoting the position of bead $i$ by $\br_i$, the
configuration of the entire network with the super-vector
$\bR=(\br_1,\ldots,\br_N)^T$ and the equilibrium
(i.e.\ minimum-energy) configuration by $\bR_{\rm eq}=(\br_{1,{\rm
    eq}},\ldots,\br_{N,{\rm eq}})^T$, the
distance between a pair of beads $i$ and $j$ by
$r_{ij}\equiv|\br_i-\br_j|$, and the stiffness and
equilibrium ``rest length'' of the bond connecting a pair of beads
by $r_{ij}^0=|\br_{i,{\rm eq}}-\br_{j,{\rm eq}}|$, respectively, the potential energy of a configuration $\bR$ is given by
\begin{equation}
U_{\rm EN}(\bR)=\frac{1}{2}\sum_{i>j}\kappa{\rm
  A}_{ij}(r_{ij}-r_{ij}^0)^2,
\label{ENM}
\end{equation}  
where ${\rm A}$ is the adjacency matrix of the underlying graph
with elements ${\rm A}_{ij}=\mathbbm{1}_{r^0_{ij}\le r_c}$ with
cut-off distance $r_c$ and $\mathbbm{1}_\chi$ being the
indicator function of the set $\chi$.
\blue{While the developed computer code allows for an adaptive,
  instantaneous configuration-dependent adjacency, 
  we keep the adjacency matrix ${\rm A}$ fixed during the entire response calculation.}~Note that the
interactions Eq.~\eqref{ENM} are manifestly an-harmonic in $\bR$.

The construction of ENM representations of proteins is well
established and straightforward.~Briefly, we retrieve the
positions and atomic assignments from the Protein Data Bank (PDB) and
extract the $\alpha$-carbons (i.e.\ the principal protein-backbone
carbon). A cutoff distance $r_c$ is chosen \blue{as the smallest
  distance at which
exactly 6 eigenvalues of the Hessian matrix of $U_{\rm ENM}$ are
zero}.~$\alpha$-carbons closer than $r_c$ are then connected with
Hookean springs, and
without much loss of generality we set the spring stiffness
to $\kappa =1$. 
\blue{The $r_c$ determined for various proteins all lie within 
$8\,\angstrom\leq r_c \leq 14\,\angstrom$ in agreement with} existing 
literature~\cite{Sanejouand_ENM_2013, kondrashov2007protein,
  kundu2002dynamics} \blue{(See Table~\ref{tab:prots} for
  details). Notably, the results are insensitive to the choice of $r_c$,
  i.e.\ a small change of $r_c$ in the above range causes no
  qualitative changes \cite{Sanejouand_ENM_2013}. We
neglect effects of rigid-body motions and
constraints imposed on the protein by the crystalline environment 
\cite{Dima_critique}. Accounting for these effects may yield more accurate specific models in terms
of 
experimental B-factors, but is not required for
our purpose as we are in pursuit of general findings.} 
In total we analyzed 14 different allosteric proteins from the PDB for which
both, the structure and allosteric binding sites are known.

In order to avoid potential
unphysical configurations during the training and response of
artificial ``pseudoproteins''---networks we design to mimic the
structure and dynamics of globular proteins---we also include the repulsive Weeks-Chandler-Andersen pair
potential between beads \emph{not} connected by a spring 
\begin{eqnarray}
U_{\rm WCA}(\bR)&=&4\epsilon \sum_{j>i}\kappa({\rm
  A}_{ij}-\delta_{i,j})\mathbbm{1}_{r_{ij}\le 2^{1/6}\sigma}\nonumber\\
&\times&\left[
  \left( \frac{\sigma}{r_{ij}}\right)^{12} -\left(\frac{\sigma}{r_{ij}}\right)^{6}+\epsilon
  \right],
\label{WCA}
\end{eqnarray}
where $\delta_{ij}$ denotes Kronecker's delta. 
In the case of protein-derived networks the inclusion of $U_{\rm WCA}$ is not required, as
no unphysical configurations occur during the response. 

The determination of binding pockets in proteins is described in
Appendix~\ref{Appendix-binding}. Moreover, the method to construct and
train artificial pseudoproteins to exhibit an allosteric response
is detailed in Appendix~\ref{Appendix-training}. The elucidation of
binding-pocket candidates in pseudoprotein
networks is described in Appendix~\ref{Appendix-pockets}.

\subsection{Full nonlinear mechanical response}
Neglecting inertial effects, which is justified by the low Reynolds
number of aqueous media and the fact that internal friction in
proteins is similar to that of water \cite{Internal}, we may determine
the response of the configuration of the network to an external force
due to ligand binding as the numerical solution of the overdamped Newton’s equations
of motion discretized in
time \cite{Flechsig_Design_2017}, or alternatively, by means of
gradient descent upon constraining the position of the subset of
source beads
\cite{nonlinear_1,nonlinear_2,Poma_Li_Theodorakis_Generalization_2018,2010_Ozkan}.    
However, as the latter are
computationally expensive for large networks, we develop an equivalent
but more efficient approach we refer to as \emph{recursive constrained
quadratic optimization}, a recursive piece-wise linear response
problem.~\blue{The method evaluates protein responses by iteratively solving a quadratic approximation of the energy function, splitting the system into constrained and free parts, yielding a full non-linear response trajectory.}

Given a set of instantaneous coordinates
$\bR^{(0)}$ (not necessarily equilibrium ones) we aim to determine the minimum energy configuration of the
network upon
constraining the subset of source beads. For small deviations
around $\bR^{(0)}$, we can expand the potential to second order
in $\bR^{(0)}$. \blue{We ensure small deviations by keeping the step size for changing 
  the constraint small enough for the quadratic approximation to remain valid, i.e.\ 
  the difference $|U_2(\bR)- U(\bR)|$ is kept within
  numerical precision.
 The quadratic approximation reads}
\begin{equation}
U_2(\bR|\bR^{(0)})\!=\!\frac{1}{2}(\bR - \bR^{(0)})^T\Hess(\bR -
\bR^{(0)})\!+\! U(\bR^{(0)}),
\label{q-form}  
\end{equation}  
where $\Hess$ is the $3N\times3N$ Hessian super-matrix with elements
$\Hess^{\alpha\beta}_{kl}=\partial_{\alpha_k}\partial_{\beta_l}U|_{\bR^{(0)}}$ for beads
$k$ and $l$ and
$\alpha,\beta=x,y,z$,
and the linear term $U(\bR^{(0)})$ does not vanish for
configurations that differ from the equilibrium configuration
$\bR^{(0)}\ne\bR_{\rm eq}$.
We set $U=U_{\rm ENM}$ for protein-derived
networks and $U=U_{\rm ENM}+U_{\rm WCA}$ for pseudoproteins. 
We split the positions $\bR$ into
constrained source beads, which we encode in the super-vector $\bC$, and
the remaining responding beads $\bS$, i.e.\
\begin{equation}
\bR\equiv \binom{\bC}{\bS}, \quad \bR^{(0)}\equiv \binom{\bC^{(0)}}{\bS^{(0)}},
\label{split}  
\end{equation}
and correspondingly also partition the instantaneous Hessian in four blocks 
\begin{equation}
  \Hess =
\begin{pmatrix}  
  \HC & \HB^T\\
  \HB & \HS
\end{pmatrix}.
\label{splitM}
\end{equation}
The quadratic form~\eqref{q-form} can now be rewritten as
\begin{equation}
  U_2(\bR|\bR^{(0)})\!=\!\frac{1}{2}\bS^T\,\HS\,\bS+([\bC-\bC^{(0)}]^T\HB^{\red{T}}-[\bS^{(0)}]^T\HS)\bS+{\rm const},
\label{q-form_2}  
\end{equation}  
where the constant term is irrelevant for the optimization
problem. For a given constraint $\bC$ the response is obtained as the
solution of $\nabla_{\bS}U_2=\mathbf{0}$, which reads
\begin{equation}
\bS=\bs^{(0)}-\HS^{-1}\HB^T(\bC-\bC^{(0)}),
\label{soln}  
\end{equation}
\blue{
  and involves the numerical inversion of the slightly smaller block $\HS$.
}
The splitting of the matrix~\eqref{splitM} is equivalent to that used
in \cite{Zheng_Brooks_Identification_2005},
and the implicit formulation of the response induced by deforming a
binding pocket~\eqref{soln} was derived in
\cite{Doniach}.
However, to determine the
full nonlinear response of the free beads to the perturbation of the
source beads, we here apply the above method recursively as follows.

\blue{For the proteins and artificial allosteric networks alike, 
				we approximate the source
        perturbation, i.e.\ the
        closing/opening of the allosteric 
				pockets, by pulling the the pocket
        beads towards (or away 
				from) their centroid (or center-of-mass) position. 
				This represents the simplest possible
        perturbation of the allosteric site
        and allows us to study the allosteric mechanism of both the holo 
				and apo structures of proteins \footnote{Envisioning a larger scale application of the presented method, this choice of perturbation is easily generalizable and also applicable to proteins without known ligand-bound structures.}.
				Curiously, we find that the predictive power of this method is 
				affected neither by the choice of the trajectory nor by 
				the specific choice of the initial configuration.
				The mechanism seems to be an inherent property of the structure, 
				as a perturbation of the source pocket always leads to a 
				significant energy uptake and subsequent release in all studied networks.}

We discretize the source perturbation into small steps yielding a
``constraint trajectory'',  $\bC(k),\,k=0,\ldots,N_{\rm step}$ and an
instantaneous Hessian at each $\bR^{(0)}=\bR(k-1)$ in Eq.~\eqref{q-form},
$\Hess_{k-1}\equiv\Hess(\bR(k-1))$, and we
solve for the free beads $\bS(k)$, yielding the full response
trajectory $\bR(k)$ via Eq.~\eqref{split} with $\bR(0)=\bR_{\rm eq}$.

\blue{To ensure accuracy,} 
the step size $\bC(k+1)-\bC(k)$ is controlled such that
  $U_2(\bR(k+1))=U(\bR(k+1))$ within numerical precision for all $k$.      
The instantaneous Hessian is determined from the instantaneous
coordinates $\bR(k-1)$ using the original rest
lengths of the springs $\{r^0_{ij}\}$. The full nonlinear response
trajectory thus reads
\begin{equation}
\mathbf{R}(k)=\inf_\mathbf{R}\, [\mathbf{R}^T\Hess_{k-1}\mathbf{R}\,\big |\,\mathbf{R}_{i\in {\rm source}}=\bC(k)],
  \label{response}
\end{equation}
and is obtained by recursively solving Eq.~\eqref{soln}. 
\blue{It is the sequence of network configurations $\bR(k)$ in
  response to adapting to the changing constraints given by
  $\bC(k)$.}~ We neglect
the possibility of multiple solution trajectories $\mathbf{R}(k)$ for
a given input $\bC(k)$; these are indeed unlikely but potentially possible in
highly symmetric networks. 
Conversely,
the linear response corresponds to the single-step solution of
Eq.~\eqref{soln} imposing the full constraint $\bC(N_{\rm
  step})$ \footnote{According to the fluctuation-dissipation theorem
we have for small $\delta\bR\equiv\bR-\bR^{(0)}$ that
$\langle\delta\bR\delta\bR^T\rangle_{\rm eq}=\Hess^{-1}_0$ and
accordingly the linear response to a small force $\mathbf{F}$,
$\delta\bR_{\mathbf{F}}=\Hess^{-1}_0\mathbf{F}$,
follows. Equivalently, one may also impose a constraint $\bC(k)$ as
we do and solve for the force on, and displacements of, the free beads
\cite{Yan_Ravasio_Brito_Wyart_Architecture_2017,2010_Ozkan}.},
i.e.\ $\bS=\bS^{(0)}-\HS_0^{-1}\HB_0^T(\bC(N_{\rm step})-\bC^{(0)})$.
The
full response to closing an allosteric pocket in the HSA protein is
illustrated in Fig.~\ref{fig:example_intro}b (green trajectory) with source (blue beads)
and target (red beads) sites which are the known binding sites of
warfarin and heme, respectively.

\section{Response to local perturbation and allosteric lever}\label{Lever}
As the initial step we construct ENMs (proteins and pseudoproteins), determine the binding-pocket 
candidates, and evaluate the full response to closing these
individually, as detailed in Sec.~\ref{setup}. Moreover, for each
configuration $\bR(k)$ during the response (see
Eq.~\eqref{response}) we determine the spectrum of the
instantaneous Hessian $\Hess_k$ 
\blue{
  i.e.\ the Hessian matrix of the potential energy $U$ evaluated at the configuration $\bR(k)$. 
}
\begin{equation}
\Hess_k\mathbf{v}_l(k)=\omega_l(k)\mathbf{v}_l(k),\quad
k=0,\ldots,N_{\rm step}\,,
\label{normal}  
\end{equation}
where
$\mathbf{v}_l(k)$ is the $l$-th (column) eigenvector (for an illustration see the first 4
nonzero eigenmodes of the HSA protein in Fig.~\ref{fig:example_intro}c)
and $\omega_l(k)$ the
corresponding eigenvalue. Eigenmodes with small $\omega_k$ are
``soft'' because their excitation requires only little
energy. Conversely,  modes with large $\omega_k$ are said to be
``stiff'' as they entail a large local curvature of $U$ and thus
display greater resistance against perturbations.
As they are irrelevant, rigid-body motions (i.e.\ zero
modes) are systematically removed in the analysis as described in Appendix~\ref{rigid}.   

 In contrast to ``simply cooperative'' elastic networks
 displaying responses which can be explained with a single
 ``dominant'' soft-mode of the unperturbed state
 \cite{Yan_Ravasio_Brito_Wyart_Principles_2018},
functional responses of proteins typically involve coordinated
motions, depending on the structure in a complex
manner~\cite{Miyashita_Onuchic_Wolynes_Nonlinear_2003,Kim_Lu_Strogatz_Bassett_Conformation_2019}. As
a result, the allosteric response generally \emph{cannot} be
rationalized in terms of soft modes alone. To visualize this, we
compare in Fig.~\ref{fig:example_intro} the full response of the HSA protein to
closing the allosteric binding pocket for the drug warfarin (panel
b, green trajectory) with the 4 softest modes of the unperturbed
network (panel c). Note that (i) the softest mode does not participate
at all in the response of the target site (red beads in
Fig.~\ref{fig:example_intro}b) and, more strikingly, \emph{none} of the four
softest modes participates in the closing of the source pocket (blue beads in
Fig.~\ref{fig:example_intro}b). That is, the perturbation and response are
orthogonal to the four softest eigenmodes of the resting network.   

Based on  (i) the fact that allostery is \emph{not} a
set of common motion patterns \cite{Gerstein,Taylor}, (ii)
is a property of all proteins that is
merely amplified in ``allosterically active'' ones
\cite{Daily_Gray_Local_2007}, (iii) functional allosteric responses
require concerted long-range motions involving multiple modes~\cite{Miyashita_Onuchic_Wolynes_Nonlinear_2003,Kim_Lu_Strogatz_Bassett_Conformation_2019},
and (iv) an optimal transmission of a structural
perturbation from the
source to the target should ``dissipate'' the minimal amount of energy
near the source pocket, 
we formulate the following
central
\begin{Definition}[Allosteric lever] Allostery is an evolutionarily
  optimized coupling pattern
  between anharmonic
modes induced by ligand binding;~the 
  perturbation of the source pocket preferentially loads stiff
  modes and relaxes along a fine-tuned
  mixture of
  softer modes that specifically rearrange the target
  pocket. Accordingly we should expect that:
  \begin{enumerate}[noitemsep,topsep=0.1pt,label=\arabic*)]
    \item Both, loading and relaxation energy are extremized. 
    \item The response is non-reciprocal and non-linear. 
    \item The true source pocket effects a \blue{distinctive} response.
    \item Allosteric transmission channels are conserved through evolution.
  \end{enumerate}
\end{Definition}

\blue{We now provide a rationale for the individual points of the
  hypothesis:}
  
\blue{\emph{1)} This point follows from the assumption
  that during evolution, the propagation
  of mechanical signals from allosteric
  to active sites was optimized for
  efficiency. ``Loading'' refers to the
  networks uptake of energy provided by
  the binding event, while
  ``relaxation'' refers to the 
  response during which the energy, taken
  up during the loading step, is
  relaxed. For an optimal transmission,
  both energy changes should be of the
  same order of magnitude, and both
  should be
  larger than most other
  possible binding events within the
  network, i.e.\ at other binding
  pockets.
}

\blue{
  \emph{2)} While non-reciprocity in allosteric systems has been previously reported \cite{hamilton2024rna,campitelli2020allostery,campitelli2018hinge}, 
  reciprocal responses have also been
  observed and even successfully
  exploited to predict allosteric sites
  \cite{Tee}. Indeed,
  some systems exhibit
  multi-directional allostery
  \cite{abhishek2023allosteric}. Our hypothesis
  extends this concept to include the
  non-linear response of the system,
  which represents a qualitative and
  quantitative advance with respect to
  previous studies. 
  Similarly, while non-linearity in allosteric systems has been previously described \cite{stock2018non, Miyashita_Onuchic_Wolynes_Nonlinear_2003}, 
  we provide here both, a mechanistic explanation for its origin and a new quantitative framework for its analysis.
  Thus, both are included in the
  hypothesis because they are important
  a central parts of the allosteric
  mechanism.
}

\blue{\emph{3)} Allosteric sites cannot be arbitrarily distributed
  throughout the protein; instead, the true source pocket must have
  distinctive properties. This is crucial for control functions within
  the protein: if every pocket were to exert allosteric effects,
  precise regulation of protein function would be impossible. Thus,
  the distinctive response of the true source pocket serves as a
  critical feature for maintaining controlled allosteric regulation.}

\blue{\emph{4)} Several studies established the conservation of both catalytic 
  \cite{fischmann1999structural,bartlett2002analysis,panchenko2004prediction}  
  and allosteric 
  \cite{clarke2016identifying}
  sites, as well as residues connecting them
  \cite{suel2003evolutionarily}.
  This conservation pattern extends to
  residues with specific mechanical
  properties, such as those with low
  flexibility \cite{campitelli2020role}
  or those acting as hinges \cite{liu2012sequence}. 
  We propose that the residues
  participating in the allosteric lever
  mechanism should similarly show
  elevated conservation, as they face
  evolutionary pressure to maintain the
  intricate coupling between hard and
  soft modes that in turn enables efficient allosteric communication.
}

The first point can be tested in a natural and straightforward
manner. According to Eq.~\eqref{response} we split the
response into two steps (in a ``Trotter'' fashion \cite{Trotter}): the
\emph{constraining step} $\delta\bC_k\equiv \bC(k)-\bc(k-1)$ and the \emph{relaxation step} $\delta\bS_k\equiv
\bS(k)-\bS(k-1)$ (see
Eq.~\eqref{split}). Accordingly, the energy take-up during loading is
given by (see Eqs.~(\ref{split}-\ref{soln}))
\begin{equation}
\delta U(\delta\bC_k)=\frac{1}{2}\delta\bC^T_k\HC_k\delta\bC_k,
  \label{takeup}
\end{equation}
\blue{involving those blocks of the hessian matrix $\HC_k$ related to the constrained beads and their displacements $\delta\bC_k$,}~while the energy release during relaxation
\blue{can be expressed as the full energy change minus the energy
  take-up during constraining and}~reads
\begin{eqnarray}
\delta
U(\delta\bS_k)&=&\frac{1}{2}\delta\bR^T_k\Hess_k\delta\bR_k-\frac{1}{2}\delta\bC^T_k\HC_k\delta\bC_k\nonumber\\
&=&\frac{1}{2}\delta\bS^T_k\HS_k\delta\bS_k+\delta\bS^T_k\HB_k\delta\bC_k\nonumber\\
&=&-\frac{1}{2}\delta\bC_k^T\HB_k\HS_k^{-1}\HB_k^T\delta\bS_k,
\label{relax}
\end{eqnarray}
where in the last line we used Eq.~\eqref{soln}. In the
  analysis we consider all steps of the response and perturb the
  pocket to a fixed percentage of its initial radius of gyration.  
  To rule out
trivial effects caused by different pocket sizes, we divide the
energy by the number of beads in the source-pocket
candidate. Moreover, to allow for a comparison of different proteins
and pseudoproteins we consider energy changes relative to that of the true
source pocket, i.e.\ $\delta U^i/\delta U^{\rm source}$. The energy
uptake and relaxation of the true source are thus $\pm 1$,
respectively.
\begin{figure*}[t!]
\includegraphics[width=1.\textwidth]{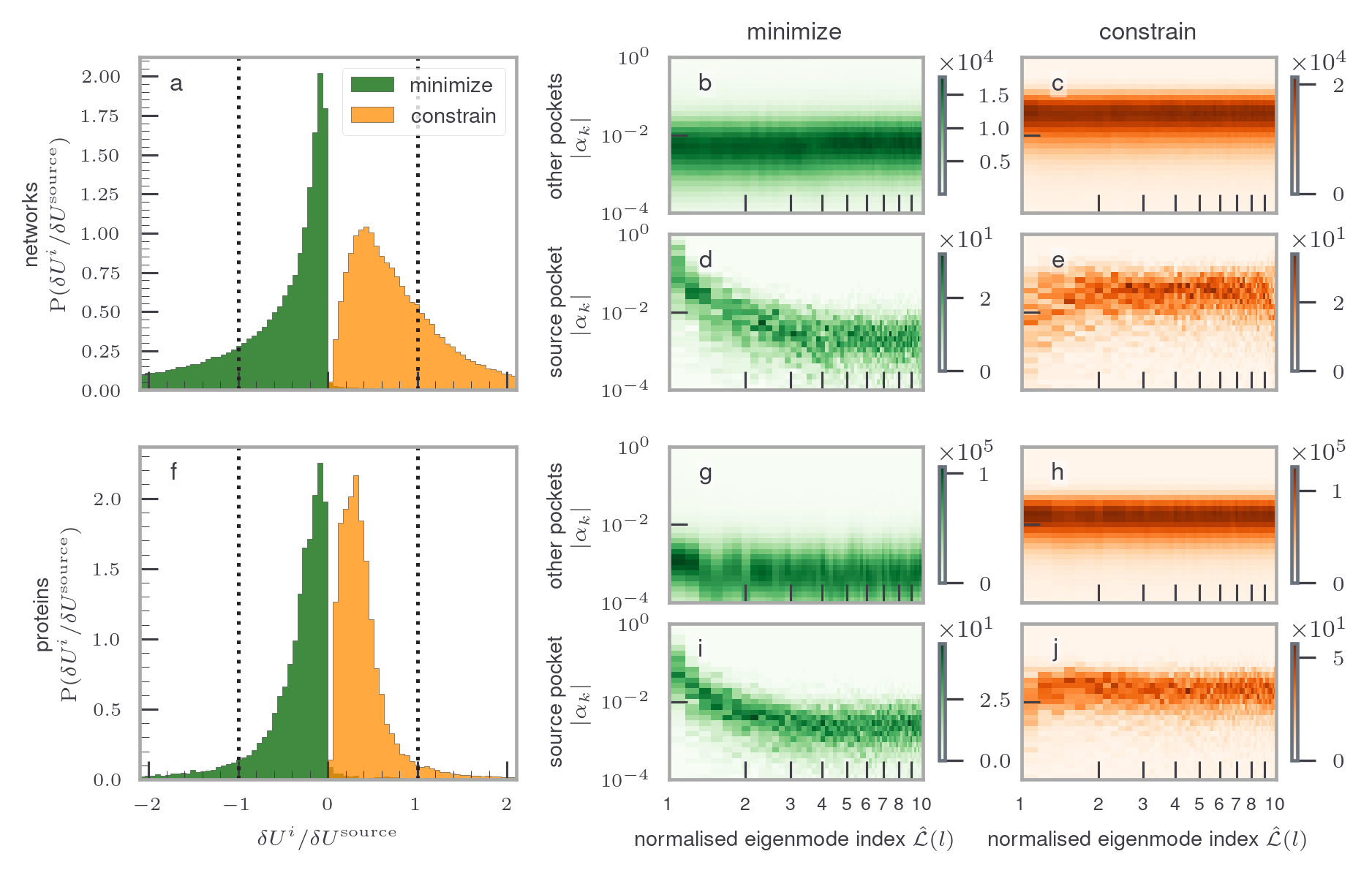}%
\caption{\textbf{Energetics of two-step response unveils the
    allosteric lever.} 
  Statistics of $\delta
  U^i/\delta U^{s}$ for all source-pocket candidates during perturbation (orange) and relaxation
(green) for pseudoproteins (\textbf{a}) and proteins (\textbf{f});
  Energy changes $\delta U$ are normalized per-bead to allow for a
  comparison of networks and proteins with different sizes. The results of the true source
  pocket are highlighted with dashed lines at $\pm1$. 
Alternatively we compare responses to perturbing
the pocket $i$ during loading  $\delta\mathbf{C}^i$ and relaxation
$\delta\mathbf{S}^i$ in terms of their projections $\alpha_k$ onto 
nonzero eigenmodes
of the corresponding Hessian matrix, $\mathbf{v}_k$
(see Eq.~\eqref{proj}) for pseudoproteins (\textbf{b-e}) and proteins (\textbf{g-j}) for true source
pockets (bottom) and other binding-pocket candidates (top); note the
logarithmic scale. To allow for a consistent comparison of structures with different sizes,
the eigenmode indices $k$ are spread evenly between 1 and 10 \red{and
  shown on the double-logarithmic scale for better visibility}; the
complete mapping onto the normalized eigenmode index is
$\hat{\mathcal{L}}(l)$ (see Eq.~\eqref{mapping}).~Both pseudoproteins and proteins show uniform
projections over the mode indices for non-allosteric pockets.~In the
case of true source pockets the perturbation does not couple to soft
modes, whereas the response in fact predominantly couples to these,
which we refer to as the \emph{allosteric lever}.~There is no
qualitative difference between pseudoproteins and proteins.}
\label{Fig2}
\end{figure*}

This analysis of the energetics of the uptake and relaxation is
illustrated for the HSA protein in Fig.~\ref{fig:example_intro}d-f, depicting
$\delta U^i/\delta U^{\rm source}$ projected on the particular
source-pocket candidate for the uptake (d) and release (f), 
while the histogram over all source-pocket candidates is shown in
Fig.~\ref{fig:example_intro}e. Only a few pockets show significant energy
changes, and the response of the true source pocket falls among the
largest energy uptake and release, respectively. Note, however, that
we find pockets producing even larger energy changes, which, however, do
not necessarily correlate with a large change at the target pocket; 
we explain these below.

To verify that the results for the HSA protein are in fact
representative, we evaluate in Fig.~\ref{Fig2} the statistics over all pseudoproteins
and all proteins. We find that the ``fat tails'' in the relaxation step
are more pronounced than in the loading step, and that both the energy
uptake and relaxation of the true source
pockets in real proteins is better optimized than in pseudoproteins (compare
panels a and f in Fig.~\ref{Fig2}). Here as well, we find that some
source-pocket candidates display even larger energy changes that
concurrently  do
\emph{not} necessarily correlate with a large response of the target
pocket. To exclude potential artifacts of sampling only subsets of the
very large binding pockets instead of the full pocket, we also analyze the energetics of the response of full
pockets (see Fig.~\ref{fig:proteins_energy_1d_hist_big_pockets} in Appendix~\ref{Appendix-binding} for details),
where we find no qualitative difference.

The outliers with large $|\delta U^i/\delta U^{\rm source}|$ that are not true source pockets
show that the relative energy change $\delta U^i/\delta U^{\rm
  source}$ alone is not a sufficient criterion. We show below that
this is because the observable $\delta U^i/\delta U^{\rm
  source}$ does \emph{not} necessarily account for the magnitude and
specificity of the response of the target pocket. In fact, we show
below (see Fig.~\ref{Fig3}g-j)
by
conditioning on the magnitude or specificity the response of the
target, $\delta U^i/\delta U^{\rm
  source}$ may in fact be used to predict allosteric source sites
based on purely physical arguments (neglecting all chemical details) 
with a quite remarkable accuracy.

To test if the observed energy changes reflect the loading of
hard modes and relaxation along softer modes upon perturbing the true
source pocket (in line with our hypothesis), we determine the
projection of the normalized loading of source pocket $i$ for each
step $k$ of the response, $\hat{\bC}_k^i=\bC^i(k)/|\bC^i(k)|$ and the
corresponding relaxation step 
$\hat{\bS}^i_k=\bS^i(k)/|\bS^i(k)|$ (both augmented to the full dimensionality
$\mathbbm{R}^{3N}$ with zeroes) onto the respective eigenvectors
\begin{equation}
\alpha_l(k)=\hat{\mathbf{x}}^T\mathbf{v}_l(k),\quad\hat{\mathbf{x}}=\hat{\bC}_k^i,\hat{\bS}^i_k\,.
\label{proj}  
\end{equation}  
To allow for a comparison of networks and proteins with different
sizes (and thus different number of eigenmodes) we divide by dimensionality
$3N$ \textcolor{black}{and map the eigenmode index $l$ for each
  network uniformly onto the interval $[1,10]$}, yielding the \emph{normalized eigenmode index}
defined as
\begin{equation}
\red{\hat{\mathcal{L}}(l)\equiv \left[9\frac{(l-6)}{(3N-6)}+1
  \right],\quad l=6,\ldots,3N.}
\label{mapping}
\end{equation}
\red{By construction $\hat{\mathcal{L}}(l)$
  lies between $1$ and $10$ for all structures.}
We inspect histograms of $|\alpha_l|$ binned as a function of
$\hat{\mathcal{L}}(l)$ for each incremental step of the response for all
pseudoproteins (Fig.~\ref{Fig2}b-e) and proteins
(Fig.~\ref{Fig2}g-j). We find that in comparison to ``false'' pockets,
the perturbation of true source
pockets evidently avoids coupling to soft modes (compare
Fig.~\ref{Fig2}c and Fig.~\ref{Fig2}e for  pseudoproteins and
Fig.~\ref{Fig2}h and Fig.~\ref{Fig2}j for proteins). Even more
prominently, the concurrent relaxation upon loading true source
pockets predominantly involves soft modes, whereas in the case
of ``false'' pockets it involves all modes equally (compare
Fig.~\ref{Fig2}b and Fig.~\ref{Fig2}d for  pseudoproteins and
Fig.~\ref{Fig2}g and Fig.~\ref{Fig2}i for proteins). This is in
line with our hypothesis and confirms \emph{point 1}.

\section{Allosteric responses are  non-linear and non-reciprocal}\label{nonreci}
\begin{figure*}[t!]
    \includegraphics[width=1\textwidth]{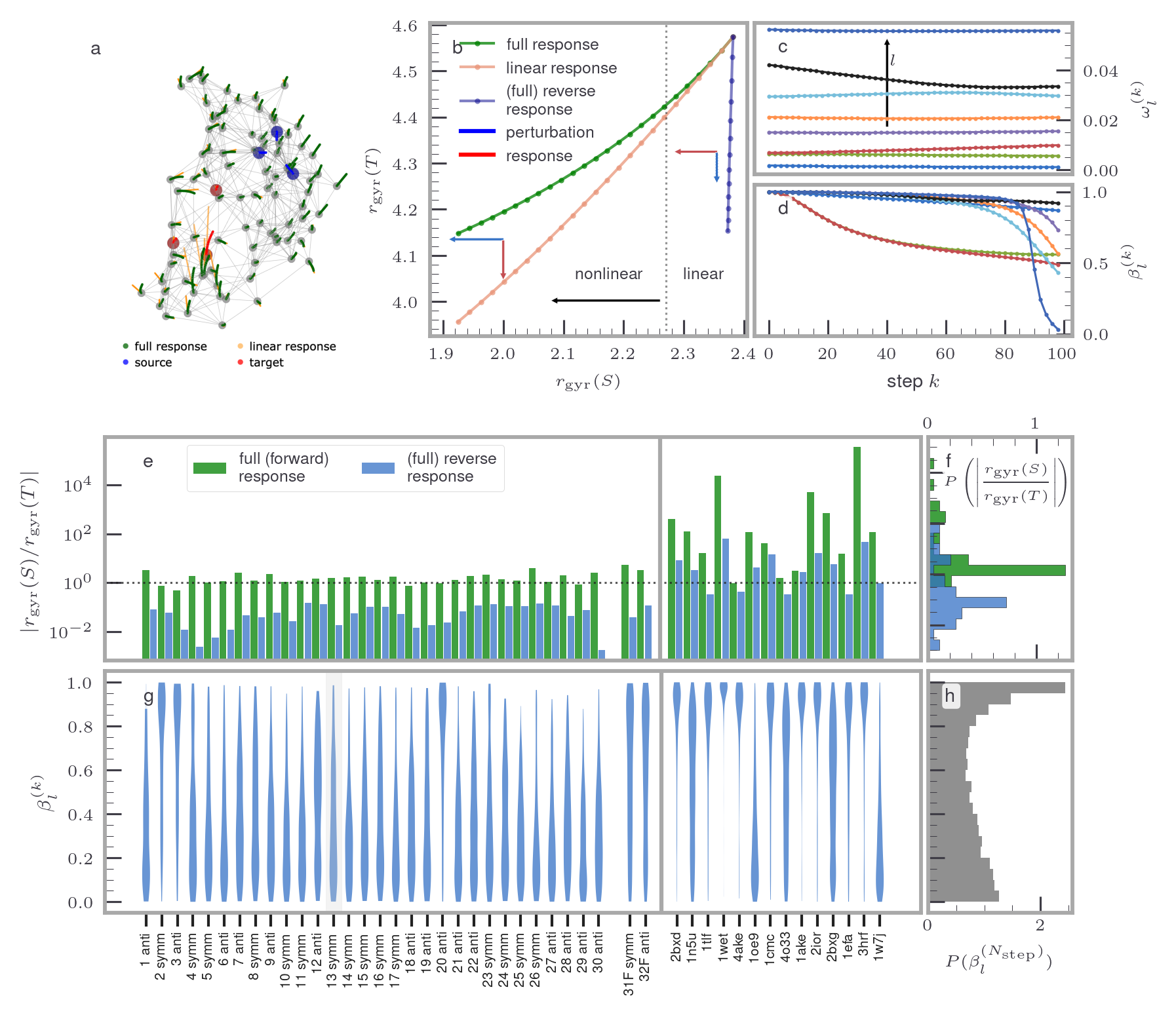}
    \caption{\label{fig:nonlin_nonreci}
    \textbf{Allosteric response is non-reciprocal and non-linear.} 
    \textbf{(a)}~Comparison of response trajectories of an example pseudoprotein determined by recursive constrained optimization (i.e.\ the
    ``full response''; green) and by a single-step linear response
    (orange) for the
    same input. The source (blue) and target (red) pockets are highlighted
    for the initial configuration.~\textbf{(b)}~Corresponding radii of gyration of the input 
    (i.e.\ source) $r_{\rm gyr}(S)$ and target $r_{\rm gyr}(T)$ 
    as quantifiers of the 
    \blue{pocket size during the full and linear
      response (lines with points in the same colors as in \textbf{(a)}) 
      and the (full) reverse response (blue line with points).}
    The arrows indicate the direction of the input (blue arrow) and output (red
    arrow) for the forward and reverse responses, respectively.
    \textbf{(c)} Changes of the the first~\blue{$9$ eigenvalues $\omega_l^{(k)}$}  
    of the ``instantaneous
    Hessian'' during the response and  \textbf{(d)} the scalar product $\beta_k^{(k)}$
    of the corresponding eigenvectors of the ``instantaneous
    Hessian'' during the respective steps of the response
    $\mathbf{v}_l(k)$ with their initial directions $\mathbf{v}_l^{(0)}$ (see Eq.~\eqref{rot}). \textbf{(e-h)}~Statistics of the results in \textbf{(a-d)} taken over all trained pseudoproteins and proteins:~\textbf{(e)}~Ratios of the input vs.\ output for the ``forward'' and
    reverse response providing clear evidence of non-reciprocity,
    histogrammed in  \textbf{(f)} (note the two clearly distinct peaks). 
    \textbf{(g)} 
    \blue{Violinplots over concatenated values of 
    the rotation (i.e.\ $\beta^{(k)}_l$) of all eigenvectors} for all
    trained pseudoproteins (including two structures designed in
    \cite{Flechsig_Design_2017}) and the 14 analyzed proteins (the
    gray 
    shaded area indicates the pseudoprotein shown in \textbf{a}-\textbf{c}). 
    Note that in many cases the initial orientation turns perpendicular
    during the response, whereas for others (those displaying a linear response) it
    remains constant throughout the response.~\textbf{(h)}~Corresponding
    histogram revealing a broad range of reorientation magnitudes during the
    response, underscoring the limited (approximate) validity of linear response theory
    only for a subset of proteins.}
\end{figure*}

We now show that, and explain why, the allosteric response to perturbing the
\emph{true} source pocket is 
typically both, non-linear and non-reciprocal. Already upon visual
comparison of the full and linear responses (see
e.g.\ Fig.~\ref{fig:nonlin_nonreci}a) we find that they typically significantly
diverge. To make the comparison quantitative, we inspect the radii of
gyration of the source $r^2_{\rm gyr}(S)$ and target $r^2_{\rm
  gyr}(T)$ pocket defined in Eq.~\eqref{rgyr} during the response (green line in Fig.~\ref{fig:nonlin_nonreci}b) and
compare them with the linear-response approximation (orange line in
Fig.~\ref{fig:nonlin_nonreci}b). Except for very small perturbation of the source,
the full and linear response disagree substantially.

Moreover, we compare the ``forward'' response to perturbing the source
pocket to that of perturbing the target pocket exactly in the same manner it
changes during the ``forward'' response. We find that the
response is strongly non-reciprocal, in line with the
allosteric-lever hypothesis, and is an immediate consequence of the
fact that multiple eigenmodes are involved. The non-reciprocity
follows directly from Eq.~\eqref{soln} which involves fundamentally
different response matrices $\HS^{-1}\HB^T$ during the forward and
inverse response and shows non-reciprocity also in the linear-response
approximation (see also Fig.~\ref{fig:nonlin_nonreci}b). That is, non-reciprocity
and non-linearity of the response are not inter-dependent. 

Intuitively, imposing
the constraint on the source beads and allowing the free beads to
relax, we confine the latter minimization to some
hypersurface. Conversely, constraining the
target beads to follow precisely their former response, the
hypersurface to which the free beads are confined during the reverse
response is obviously different, and thus so will be the
minimum-energy path. Therefore, the ``forward'' and reverse trajectories of the respective free beads
will generally \emph{not} be the same. Note that there is conclusive recent
evidence for a one-way allosteric communication obtained by
atomisctic Molecular Dynamics simulations
\cite{Graeter_oneway}. \blue{For experimental evidence of a
  bi-directional, but \emph{not} symmetric, allosteric control via photoswitching
  see \cite{Hamm}.}
However, the situation is
different when the response goes along a single soft mode \cite{Yan_Ravasio_Brito_Wyart_Principles_2018}, since here the motion
is effectively one-dimensional, such that the free beads during relaxation
are confined to the same line and the response is thus manifestly
reciprocal.

Hints about the origin of the non-linearity of the response in turn come
from the observation of
avoided crossings in the behavior of the eigenvalues of the
instantaneous Hessian matrix $\Hess_k$ during the full response (see
Fig.~\ref{fig:nonlin_nonreci}c) and substantial rotations of the corresponding
eigenvectors quantified via the projection of the instantaneous
eigenvectors $\bv_l(k)$ at step $k$ of the response onto those of the resting Hessian $\bv_l(0)$
(see Eq.~\eqref{normal}),
i.e.\
\begin{equation}
\beta^{(k)}_l\equiv \bv_l(0)^T\bv_l(k),\quad k=0,\ldots,N_{\rm step},
  \label{rot}
\end{equation}
which show large distortions of the eigendirections of normal modes
during the response (see Fig.~\ref{fig:nonlin_nonreci}d).  
This rationalizes the failure of linear response theory but does not
yet explain its origin.

Before explaining this deeper, we first check if the
non-linearity and non-reciprocity of the allosteric response are a
\emph{typical} observation in both, pseudoproteins and proteins. 
Indeed, a comparison of the magnitude of the relative change at the
source versus target 
pocket $|r^2_{\rm gyr}(S)/r^2_{\rm gyr}(S)|$ (see Fig.~\ref{fig:nonlin_nonreci}e)
clearly confirms that the ``forward'' and reversed responses are
almost always asymmetric, which is quantitatively shown in the
frequency  histograms of $|r^2_{\rm gyr}(S)/r^2_{\rm gyr}(T)|$ in
Fig.~\ref{fig:nonlin_nonreci}f.

Similarly, the projection of the instantaneous onto the resting
eigenvectors encoded in $\beta^{(k)}_l$ typically significantly
differs from 1, indicating the failure of linear response (see
Fig.~\ref{fig:nonlin_nonreci}g), which is quantified in 
frequency histograms of $\beta^{(k)}_l$ depicted in
Fig.~\ref{fig:nonlin_nonreci}h. Taken together, this 
strengthens the
evidence supporting the allosteric-lever hypothesis and in particular
it confirms  \emph{point 2}.

\subsection{Avoided eigenvalue crossings and origin of non-linear response}\label{avoid}
\begin{figure*}[t!]
\includegraphics[width=1.\textwidth]{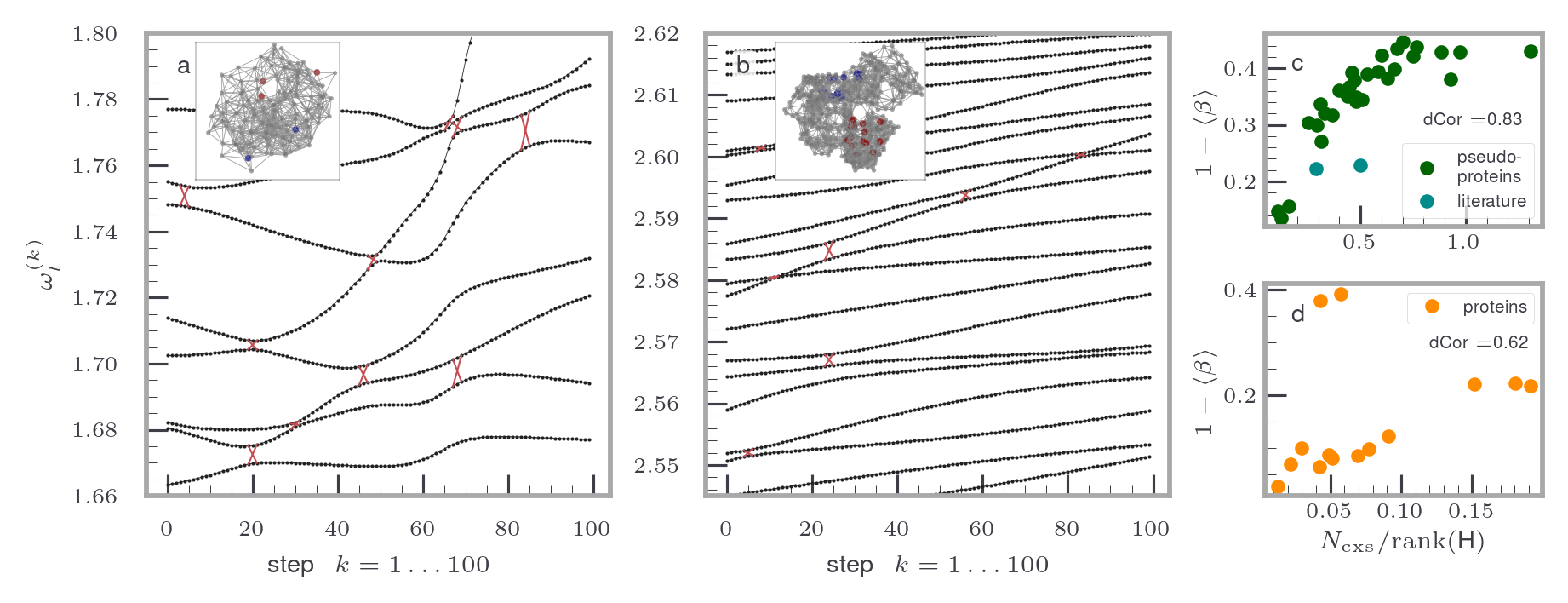}%
\caption{\textbf{Nonlinear response and avoided crossings.}
\textbf{(a-b)}~Eigenvalues
  $\omega_l(k)$
  of the instantaneous Hessian at the $k$-th step of the response of \textbf{(a)}
  a self-trained pseudoprotein and \textbf{(b)}~HSA protein (PDB ID: 2bxg~\cite{ghuman2005structural});~red crosses indicate
  the estimated positions of avoided crossings 
  and the 3D insets show
  representations of the respective
  network.~\textbf{(c-d)} Scatter
  plots depicting a clear correlation between the number of avoided crossings
  (normalized by the total number of internal eigenvectors), $N_{\rm cxs}/{\rm rank( \Hess )}$,  
  and the nonlinearity of the response
  quantified as the rotation of
  the eigenvectors of the
  instantaneous Hessian   
  during the response with
  respect to the initial
  (resting) configuration
  averaged over the response trajectory,
  $\langle \beta \rangle$ in
  Eq.~(16), for pseudoproteins
  (\textbf{c}) and proteins
  (\textbf{d}). The two highly
  nonlinear outliers in
  \textbf{(d)} (both structures
  of the same motor protein
  (Myosin V), performing larger
  motions) are discussed in
Appendix~\ref{Appendix_crossings_end}.~\red{The magnitude of the
  nonlinear correlation is quantified in terms of the distance
  correlation ${\rm dCor}$ defined in Eq.~\eqref{distcor}.}
}
\label{fig:crossings}
\end{figure*}

In Fig.~\ref{fig:nonlin_nonreci}c we observed avoided crossings of the eigenvalues
of the instantaneous Hessian $\omega_l(k)$ (see Eq.~\eqref{normal}) as
a result of perturbing the source pocket. We now analyze these
systematically. We first determine
mid-point tangents $T_l(k)$ for all points $k$ along the eigenvalue
curves, excluding the first and last point of each curve:
\begin{equation}
T_l(k)=\frac{1}{2}[\omega_l(k+1)-\omega_l(k-1)],\,k=2,\ldots,N_{\rm step}-1, 
\label{tagent}  
\end{equation}  
and determine the points where nearest-neighbor tangents (i.e.\ $l$
and $l+1$) would intersect. An avoided
crossing between a given pair $l$
and $l+1$ occurs when the intersection point
first diverges towards $+\infty$ as a function of $k$ and subsequently
re-enters from $-\infty$ (see Fig.~\ref{avoided} in Appendix~\ref{Appendix_avoided}). Within the interval
of the ``discontinuity'' of the intersection point, the actual
location of an intersection is determined on the basis of a
sign-change of $\omega_{l+1}(k)-\omega_{l}(k)$. Only instances where the
crossing occurs first from above ($+\to -$) and then from below ($-\to +$)
are considered. Examples of avoided eigenvalue crossings determined
this way for for an artificial pseudoprotein and the
HSA protein (PDB ID 2bxg~\cite{ghuman2005structural})
are shown in Fig.~\ref{fig:crossings}a and Fig.~\ref{fig:crossings}b,
respectively.

We quantify the degree of non-linearity of the response in terms of
\blue{
    the average rotation of the eigenvectors along the response trajectory, i.e.\
  \begin{equation}
    \langle \beta \rangle = \frac{1}{3N-6}\sum_{l=6}^{3N} \frac{1}{N_{\rm step}} \sum_{k=1}^{N_{\rm step}}\beta^{(k)}_l.
    \label{beta}
  \end{equation}
  In addition to averaging over the entire trajectory as in
  Eq.~\eqref{beta}, we determine in Appendix~\ref{Appendix_crossings_end} the average rotation of the eigenvectors  
  only between the initial configuration and the end of the response
  to inspect whether the non-linearity persists or changes throughout the
  response. We find no significant differences between
  the two approaches.}

We observe a strong \red{(not necessarily linear)} correlation between 
the degree of non-linearity and 
  the number of avoided crossings $N_{\rm cxs}$ relative 
  to the rank of the matrix $\Hess$ (see Fig.~\ref{fig:crossings}c and
  \ref{fig:crossings}d for pseudoproteins and proteins, respectively;
  the two highly non-linear ``outliers'' in
  Fig.~\ref{fig:crossings}c are discussed in
  Appendix~\ref{Appendix_crossings_end}).~\red{We quantify this
 (generally nonlinear) correlation by means of the \emph{(sample) distance correlation}, ${\rm
   dCor}$ \cite{dist_cov}. That is, we consider the tuples $V_1\equiv
    \{\frac{N_{\text{cxs}}}{\text{rank}(\mathbf{H})}(i)\}_{i=1,n_s}$ and $V_2 \equiv \{ 1
    - \langle\beta\rangle(i)\}_{i=1,n_s}$ where $n_s$ denotes, respectively, the
    number of protein and pseudoprotein structures considered in the
    analysis, and define pairwise distances $D^\alpha_{kl}\equiv
    |V_\alpha(k)-V_\alpha(l)|$ for $\alpha=1,2$ in the Euclidean norm
    $|\cdot |$. We further define
     the (doubly centered) symmetric distance matrices $\M_\alpha$ with elements
    \begin{equation}
(\M_\alpha)_{ij}=D^\alpha_{ij}-\frac{1}{n_s}\sum_{k=1}^{n_s}(D^\alpha_{ik}+D^\alpha_{kj})+\frac{1}{n_s^2}\sum_{k,l=1}^{n_s}D^\alpha_{kl},
    \end{equation}  
    as well as (scalar-valued) distance covariances
    \begin{equation}
      {\rm dCov}(V_1, V_2)\equiv \frac{1}{n_s^2}{\rm Tr}\,\left( \M_1\M_2^T\right)
          \end{equation}   
      and distance variances $\text{dVar}(V_\alpha)={\rm
        dCov}(V_\alpha, V_\alpha)$ that enter the definition of the distance correlation
    \begin{equation}
{\rm dCor}(V_1, V_2) \equiv \frac{{\rm dCov}(V_1, V_2)}{\sqrt{\text{dVar}(V_1) \text{dVar}(V_2)}}.
      \label{distcor}
    \end{equation}
  We have $0\le {\rm dCor}(V_1, V_2)\le 1$ and ${\rm dCor}(V_1,
  V_2)=0$ if and only if $V_1$ and $V_2$ are independent
  \cite{dist_cov}, so ${\rm dCor}(V_1, V_2)>0$ indeed quantifies
  statistical dependence between $V_1$ and $V_2$.~We find ${\rm dCor}\simeq 0.62$ for the
  proteins and  ${\rm dCor}\simeq 0.83$ for pseudoproteins,
  respectively, quantitatively confirming substantial correlations between the degree of non-linearity and 
  the number of avoided crossings $N_{\rm cxs}$.}

  The observed correlation is \emph{not} surprising; avoided crossings are known to occur when 
vibrations (in classical as well as quantum systems
\cite{Marcus_Ann,Marcus_Rev,Marcus_JCP})  
become strongly coupled by an external perturbation.

In simple terms, it is known that there exist two fundamentally different
regimes in the classical mechanics of coupled oscillators,
\emph{regular} (or quasiperiodic) and \emph{irregular} (or
chaotic/ergodic). The regular regime is similar to the motion of uncoupled oscillators,
i.e.\ trajectories are confined to only a limited number of the
energetically allowed disjoint regions (effectively the set of
non-communicating trajectories of a set of uncoupled
oscillators)~\cite{Marcus_Ann,Marcus_Rev,Marcus_JCP,CHIRIKOV}. 
Conversely, in the irregular regime trajectories occupy all (or almost all) energetically
allowed regions \cite{Marcus_Ann,Marcus_Rev,Marcus_JCP,CHIRIKOV}.~The
transition from (predominantly) regular to (predominantly) irregular
motion occurs once enough energy is injected to the system
or when the coupling between oscillators becomes strong enough,
and is accompanied by avoided crossings in linearized spectra
signaling an-harmonic resonances \cite{Marcus_Ann,CHIRIKOV}.~These
produce extensive  
changes in the eigenmodes and in the energy distribution among participating oscillators
\cite{Marcus_JCP,CHIRIKOV}. 

Applied to our allostery setting, a small perturbation of the source pocket
(i.e.\ within the ``regular'' regime) only excites/deforms individual eigenmodes
of the resting network, but does not couple distinct eigenmodes. 
In this regime the response is manifestly linear (see small-perturbation
regime in Fig.~\ref{fig:nonlin_nonreci}a-b) and thus the ``allosteric lever''
does not yet properly unfold as there is only a very weak or even no mode
coupling. Conversely, as the 
perturbation of the source pocket becomes stronger, modes begin to
couple substantially, which is manifested in avoided crossings. Notably, it couples
many modes (see Fig.~\ref{fig:crossings}a-b) and thus allows to transmit energy
from the hard modes excited at the source pocket to soft modes in the rest of the
network and in particular to the target pocket as observed in
Fig.~\ref{Fig2}. This gives rise to a non-linear response and is also
the mechanism of the hypothesized ``allosteric lever''.

\section{Specificity of allosteric responses}\label{sec_specific}
As announced in Sec.~\ref{Lever}, the energy changes during loading $\delta
U(\delta\bC)$ and $\delta
U(\delta\bS)$
in Eqs.~(\ref{takeup}-\ref{relax}) alone do not completely capture the
essence of the allosteric-lever hypothesis, as they do not
necessarily account for the magnitude and specificity of the response
induced in the target pocket. Note, moreover, that a precise, specific rearrangement of the
active target site is believed to be required
\cite{Daily_Gray_Local_2007}. We will show that by
comparing  $\delta
U(\delta\bC)$ and $\delta
U(\delta\bS)$ for source-pocket candidates \emph{conditioned on the magnitude and specificity of the
target response}, we can in fact characterize and predict allosteric source sites in
proteins remarkably accurately.  

Let $\mathbf{T}^0\in \mathbbm{R}^{3N_T}$ denote the initial 
configuration of the target pocket comprised of $N_T$ beads,
and $\mathbf{T}_{i}^{\rm fin}\in \mathbbm{R}^{3N_T}$
and $\mathbf{T}_{*}^{\rm fin}\in \mathbbm{R}^{3N_T}$ the final
response of the target upon perturbing the $i$-th surface pocket
and the true/optimal source pocket, respectively. In drug design
applications $\mathbf{T}_{*}^{\rm fin}\in \mathbbm{R}^{3N_T}$
would be the \emph{desired response}. Note that only the response
at the target site matters, the response of the rest of the network is
irrelevant for drug design applications. 
We therefore apply an optimal roto-translation \cite{Kabsch} to
best overlay the respective target configurations.  
We introduce the relative
magnitude of the response to closing the $i$-th source-pocket candidate as
\begin{equation}
  \Delta_i\equiv \frac{|\mathbf{T}_{i}^{\rm fin}-
  \mathbf{T}^0|}{|\mathbf{T}_{*}^{\rm fin}- \mathbf{T}^0|},
\label{relative}  
\end{equation}
and the relative distance from the desired response as
\begin{equation}
D_i\equiv\frac{|\mathbf{T}_{i}^{\rm fin}- \mathbf{T}_{*}^{\rm fin}|}{|\mathbf{T}_{*}^{\rm fin}- \mathbf{T}^0|}.
\label{desired}  
\end{equation}
For convenience we introduce the \emph{response specificity}
as
\begin{equation}
0\le \mu_i\equiv (1+D_i)^{-1}\le 1
\label{specific}  
\end{equation}
such that the desired/optimal response has relative magnitude and specificity
one, $\Delta_*=\mu_*=1$. Triangle inequalities between the
norms in Eqs.~(\ref{relative}-\ref{desired}) (see \footnote{We have,
setting $\mathbf{a}\equiv\mathbf{T}_{i}^{\rm fin}-
  \mathbf{T}^0$, $\mathbf{b}\equiv\mathbf{T}_{i}^{\rm fin}- \mathbf{T}_{*}^{\rm
  fin}$ and $\mathbf{c}\equiv \mathbf{T}_{*}^{\rm fin}- \mathbf{T}^0$,
  that $\mathbf{a}=\mathbf{b}+\mathbf{c}$ and hence $|\mathbf{a}|\le
  |\mathbf{b}|+|\mathbf{c}|$, $|\mathbf{c}|\ge
  ||\mathbf{b}|-|\mathbf{a}||$ and $|\mathbf{b}|\ge
  ||\mathbf{c}|-|\mathbf{a}||$;~the squeeze bound Eq.~\eqref{triangle} follows. Note also that rotations and translations preserve metric
properties \cite{Fogolari}.} for details) confine all responses to
the domain
\begin{equation}
\frac{1}{2+\Delta_i}\le \mu_i\le \frac{\theta(1-\Delta_i)}{2-\Delta_i}+
\frac{\theta(\Delta_i-1)}{\Delta_i},
\label{triangle}
\end{equation}
where $\theta(x)$ is the Heaviside function
being $1$ when $x>0$, $1/2$ when $x=0$, and
$0$ otherwise. 

We assume now that the target site is known. The aim is to allosterically control and
rearrange this target in a specific, desired manner in order to modulate the
binding of ligands.  
The allosteric source pocket is in turn  unknown
and to be determined.
Note that enforcing the desired response at the target site and simply
observing the rest of the
network to deduce the allosteric site is \emph{not} possible, because
the response was found to be non-reciprocal, see
Sec.~\ref{nonreci}. Therefore, whereas this
might work in rare reciprocal systems (see e.g.\ \cite{Tee,Ni}), discovering
allosteric sites by assuming a reciprocal bidirectional coupling will
generally fail.

\begin{figure*}[t!]
\includegraphics[width=1.\textwidth]{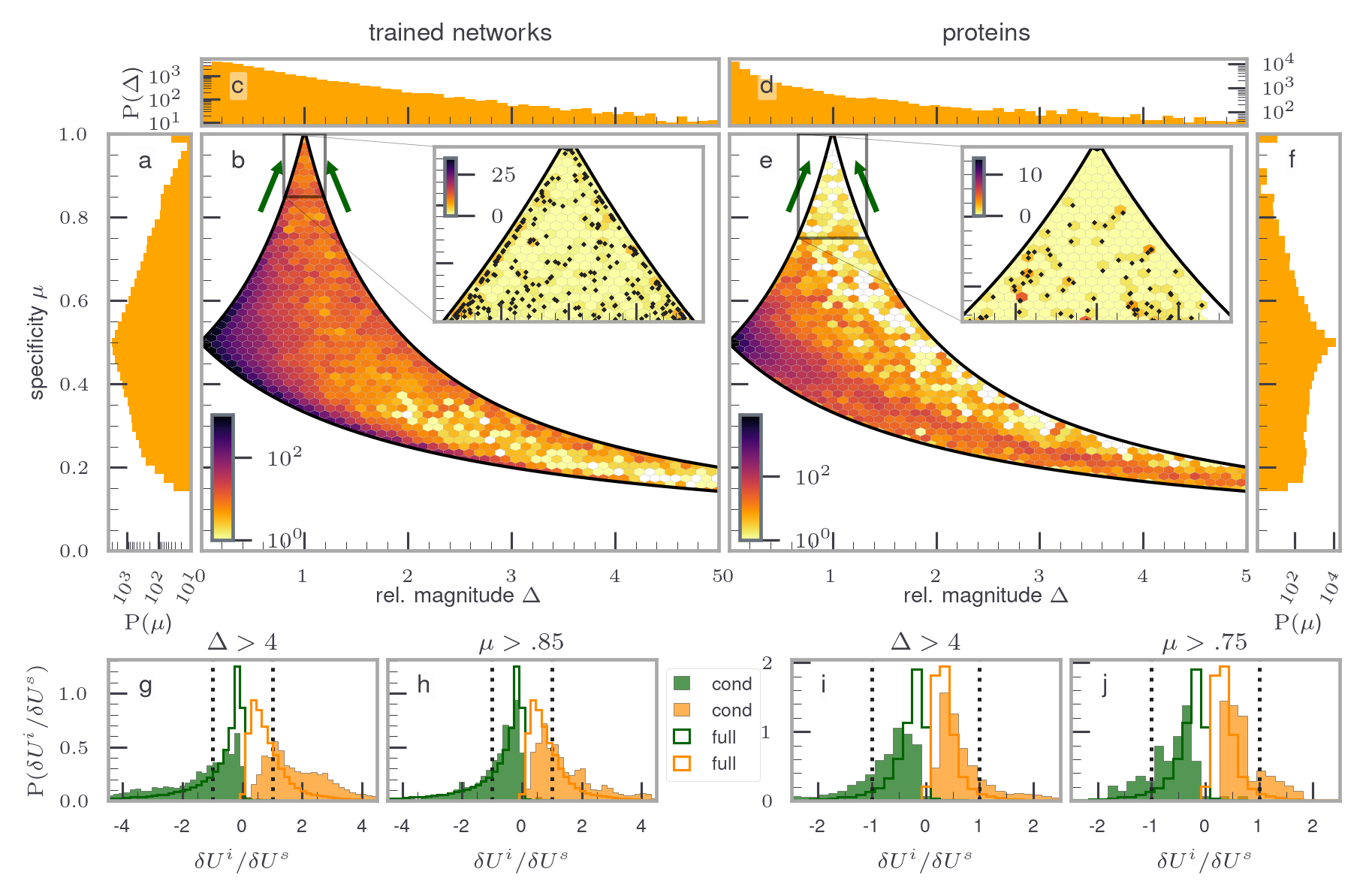}%
\caption{\textbf{Optimal source pockets efficiently transmit energy
    and yield large and specific allosteric
    responses.} Joint frequency histograms of magnitude $\Delta$ and
  specificity $\mu$, $p(\Delta,\mu)$
   (defined in Eqs.~\eqref{relative} and \eqref{specific},
  respectively) and their marginals, $p(\Delta)$ and $p(\mu)$, 
  for adjacent pairs of beads on the surface
  of pseudoproteins \textbf{(a)} and proteins \textbf{(b)}.
  The histograms are evaluated for $\Delta_i$ and $\mu_i$ for all
  pairs $i$
  of surface beads normalized with respect
  to the given network, and the statistics is determined over all
  networks. The black lines denote the bounds 
  \textcolor{red}{Eq.~\eqref{triangle}} imposed by the three
  triangle inequalities between the
norms in Eqs.~(\ref{relative}-\ref{desired}).   
         Green arrows point to the ideal response at
         $(1,1)$. The latter may not necessarily be reached in
         practical applications (e.g.\ in drug design); in this case source candidates
         nearest to the optimum are to be considered.
       Inset: magnification of the desired region. \textbf{(g-j)}
       Probability densities of energy changes $\delta U^i$ upon
       perturbing site $i$
       relative to that of perturbing the true source $\delta U^s$  during
       loading (orange shaded) and relaxation (green shaded) substeps
       conditioned on a large response
       $\Delta$ (panels \textbf{(g)} and \textbf{(i)}) and a highly
       specific response $\mu$  (panels \textbf{(h)} and \textbf{(j)}); green and orange lines correspond to
       unconditional probability densities from Fig.~\ref{Fig2} and unveil
       a clear shift towards the tails. The response was divided in 100
       increments for both, pseudoproteins and proteins.}
\label{Fig3}
\end{figure*}

We scan the response of all possible pocket pairs and triplets, for proteins also
larger pockets (see Eq.~\eqref{response}). Nearest neighbors of the target pocket are
omitted from the scan. The computing time for all 14 proteins and 33 pseudoproteins
is only a few 
hours 
on a medium-sized compute cluster. 
The exact number of
scanned pocket candidates is given in Tables \ref{tab:prots} and
\ref{tab:nets}. 
The joint
probability density of magnitude and specificity, $p(\Delta,\mu)$, and both
marginals, $p(\Delta)$ and $p(\mu)$,
over all pseudoproteins and proteins are shown in Fig.~\ref{Fig3}. 

Most candidate pockets show almost no response at all, $\Delta\approx
0$, these are located at $\mu\approx 1/2$. This is in stark contrast to the
observations on networks designed to propagate simple displacements \cite{Yan_Ravasio_Brito_Wyart_Principles_2018}, where
a large response at the target site could be triggered by perturbing
sites anywhere in the system. It is, however, in line with what is
expected for proteins. Interestingly, we also observe a considerable amount
of source pockets that yield a large response, i.e.\ $\Delta\gg 0$,
which is, however, \emph{not} in the desired direction, that is
$\mu<1/2$. These are only of secondary interest for drug design. 

Only very few candidate pockets yield both, a large
\emph{and} specific response with $\Delta\approx 1$ and $\mu\approx
1$ (see inset of Fig.~\ref{Fig3}b and e). Notably, pseudoproteins and 
proteins display qualitatively similar behavior, except that in proteins fewer
pocket candidates yield a response close to the desired one $\mu\to
1$, possibly due to a stronger specificity as a result of a superior
training by natural evolution with additional constraints that were not
accounted for in the training of our pseudoproteins. 
True allosteric source pockets in proteins thus yield remarkably
specific responses, confirming \emph{point 3}
of the allosteric-lever hypothesis.  

In drug-design applications a ``perfect'' ($\mu=1$) desired response will likely
not be reached. In this case source candidates closest to the optimum
have to be considered.

We now turn back to the energy changes during loading $\delta
U(\delta\bC)$ and relaxation $\delta U(\delta\bS)$ in
Eqs.~(\ref{takeup}-\ref{relax}) but now condition these on either a
large or specific response at the target site. 
The conditional histograms for the loading
and relaxation step are shown in Fig.~\ref{Fig3}g-j for 
pseudoproteins (g-h) and proteins (i-j), respectively. We find a clear
shift toward higher absolute values of $\delta U$ as expected, 
confirming that these efficient-loading-relaxation pockets
indeed propagate allosteric signals specifically to the target site.

\section{Predictive power of one-step relaxation energy}\label{sec_predictive}
The evaluation of the complete response and conditioning on a large
$\Delta$ or $\mu$ in principle already offers a powerful method. It
is, however, still too expensive for large-scale screening
applications in drug design. We now demonstrate that by rating the
source-pocket candidates according to the energy changes during loading $\delta
U(\delta\bC)$ and relaxation $\delta U(\delta\bS)$, respectively, already \emph{in a
single-step} response, that is, without evaluating the full response,
yields 
a practically applicable method for discovering
allosteric source sites. 
\begin{figure*}[t!]
\includegraphics[width=1.\textwidth]{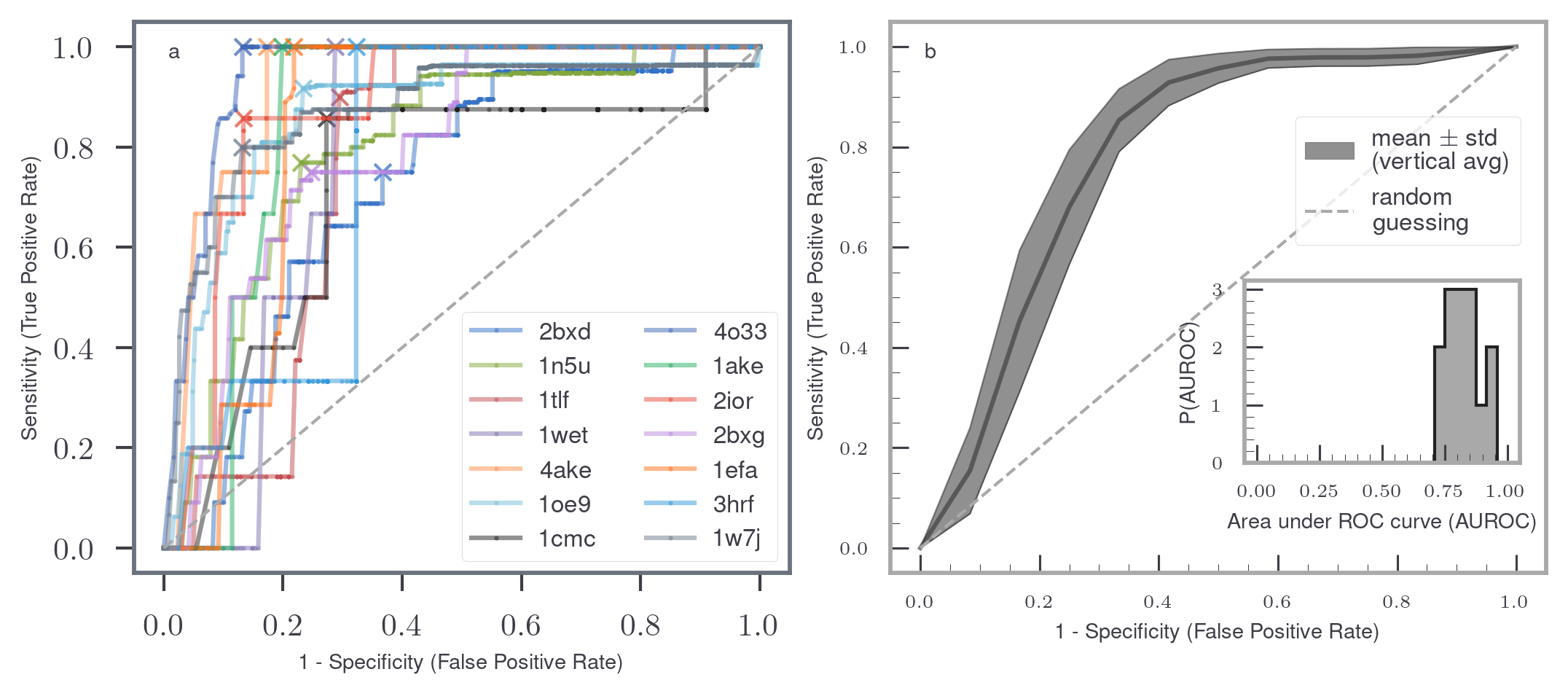}%
\caption{\textbf{Predictive power of the ``allosteric lever''.} (\textbf{a})
  Receiver-operating-characteristic (ROC)~\cite{park2004receiver}
  curves for all considered proteins highlight a strong predictive
  performance achieved by ranking candidate beads by the amount of
  strain-energy relaxation (see SM for equivalent analysis based on
  strain-energy increase during loading). Crosses depict the optimal
  threshold per protein. The gray line and shaded area indicate the 
  mean $\pm$ standard deviation of individual ROC curves obtained 
  by vertical averaging that treats each ROC curve as a function sampled 
  at fixed False Positive Rates and averaged over the True Positive Rates~\cite{fawcett2006introduction}. 
  	Already at a small
    (i.e.\ $~0.3$) False Positive Rate the True Positive Rate
    essentially converges to 1, rendering the procedure highly
    suitable for drug screening applications where it is essential to \emph{not} miss
    targets.
  The inset shows the histogram for the 
  areas under the individual proteins'
  ROC curves (AUROC).
}
\label{fig_predict}
\end{figure*}

We rank source-pocket candidates according to how much energy
the network releases during a single relaxation step $\delta
U(\delta\bS)$.  The results obtained by using instead the energy
uptake during loading $\delta
U(\delta\bC)$ are similar on average, but show a larger variance (see
Fig~\ref{fig:ROC_constrain} in Appendix~\ref{app:ROC_constrain}). 
As a large relaxation energy already requires a strong loading during
the constraining step, it effectively already includes information
about both sub-steps. Therefore, we expect relaxation energies to be more relevant.

Our data, however, is on a pair or triplet (with a few exceptions also
quartet and quintets) bead basis. There is no unique way to
reduce this data to a ``per bead'' basis. Here we use the simplest mapping: We rank pocket
candidates based on the results entering Fig.~\ref{Fig2} (but
considering a single loading-relaxation step only,
i.e.\ linear-response screening) and append
for each of the pockets its respective beads as an ordered list.
If a bead is already in the list, it is not appended again. This permits a complete ranking of all the sampled beads.
According to our prediction the beads at the top of the list are likely to be source
beads.

To generate a binary assignment, we require a threshold that classifies
beads into source and non-source beads. For a Receiver Operating
Characteristic (ROC) analysis \cite{park2004receiver}, this threshold
is traversed, and the predictions are evaluated subsequently. They fall into the classes of \emph{true} and \emph{false positives}, which are
exactly the coordinates of the ROC diagram. 
A perfect classifier would give only true positives and zero false
positives, i.e.\ a step function in the ROC graph. Conversely, a
random selection would yield a straight line with slope one half. 
Classifiers lying above this straight line are better than a random guess, and the closer
the ROC curve is to the step function (which is easily measured in terms of the
Area Under Receiver Operating Characteristic curve (AUROC)), the
better the classifier performs. 
The prediction of source beads based on $\delta
U(\delta\bS)$ performs remarkably well, as is shown in Fig.~\ref{fig_predict}.

The ROC curves for all proteins are substantially better than a random guess
and reach high values for the AUROC values, with all of them being larger than 0.7. Our
method in particular significantly outperforms a recent study with a similar aim, which assumed
the reciprocity of allosteric signal propagation \cite{Tee}. This may appear a
bit surprising, as remarkably little information enters our analysis; only a single step
of the linear-response algorithm is needed and in
particular the full response is not accounted for. It certainly
further underscores that allosteric 
responses are \emph{not} reciprocal, which seems to explain why our method
is superior already at the linear-response level. More elaborate full
responses and conditioning on specificity will further improve the
predictive power, but it is not clear if these are even required for
screening applications.~Of course, the method is only suitable for a
first screening of allosteric sites, as afterwards one must
necessarily also
take into account chemical details (charge distributions etc.)
to dock a particular ligand.

\section{Evolutionary conservation of allosteric transmission channels}\label{evo}
We now come to the final, fourth point of the allosteric-lever
hypothesis, i.e.\ that allosteric transmission channels are expected
to be conserved
through evolution.~Notably, energetic-connectivity pathways are an inherent
property of the proteins' tertiary
structure and do \emph{not} depend on secondary
structure \cite{pathways}.~Thus, the analysis based on ENMs is
expected to provide a representative qualitative picture of allostery in
real proteins as long as specific effects (e.g.\ dehydration
\cite{Schirmer_Evans_Structural_1990}) are not essential.

In evolutionary biology conserved sequences refer to identical or
similar sequences found 
in proteins or nucleic acids (RNA and DNA) across different species. A
conserved sequence pattern
is selectively maintained through evolution; the
interpretation is that the sequence pattern is functionally relevant for the protein. This
interpretation, however, has to be considered with caution, as there exist also non-coding sequences in DNA that are conserved, at first sight without
functional relevance \cite{Asthana}.
\begin{figure*}[t!]
  \includegraphics[width=1.\textwidth]{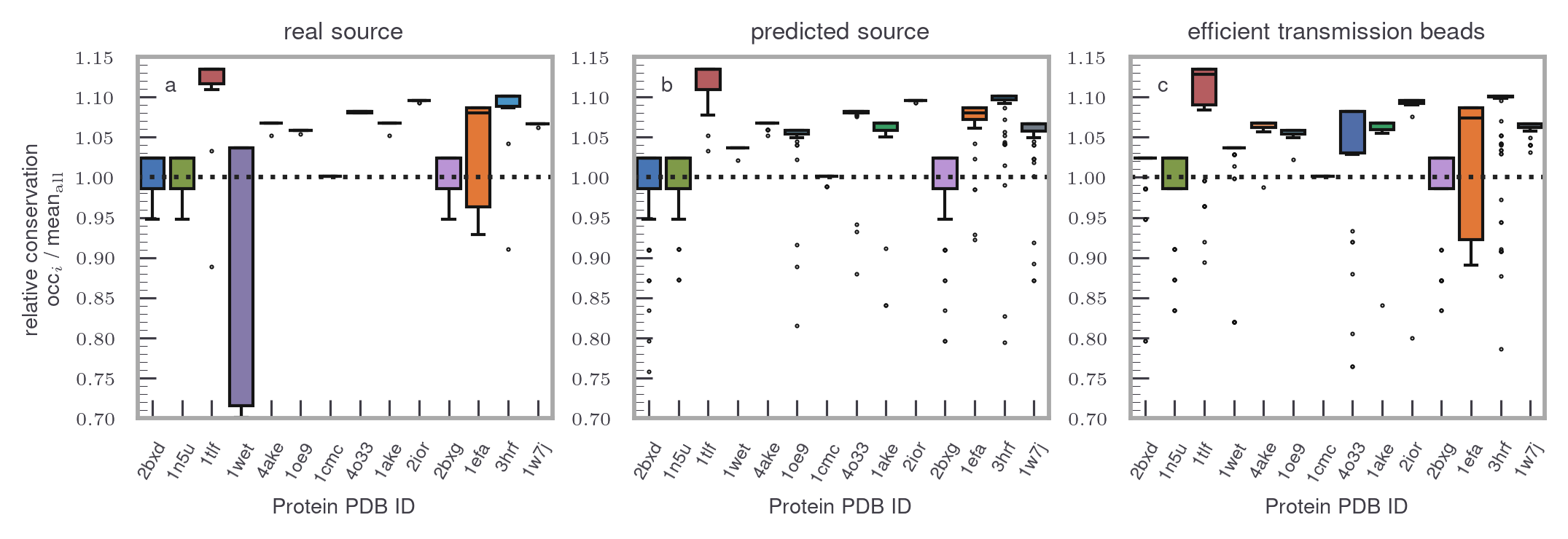}%
  \caption{\textbf{Evolutionary conservation of participating residues.}\textbf{(a-c)}
    Boxplots of statistics over the mean degree of conservation per type
    of beads for all proteins for: \textbf{(a)} true allosteric sites,
    \textbf{(b)} predicted allosteric sites, 
    and \textbf{(c)} those participating dominantly in the energy transmission. 
    Efficient transmission beads are those connected by a spring that
    contributes strongly to the energy change; we determine these beads
    by ranking the springs according to their contribution to the energy
    release during the final relaxation step.
    Note that there is less variation in the conservation of predicted sequences.}
  \label{fig:conservation}
  \end{figure*}
\blue{
  Catalytic sites are known to be highly conserved across homologous proteins, though this conservation 
  varies with functional similarity \cite{fischmann1999structural,bartlett2002analysis,panchenko2004prediction}. 
  Similarly, allosteric sites show significant conservation 
  compared to background residues \cite{clarke2016identifying}, as do the residues forming communication pathways between these 
  sites \cite{suel2003evolutionarily,liu2012sequence,campitelli2020role}. It is conceivable that this reflects the interconnected nature of protein function, where evolutionary 
  pressure maintains all three elements while allowing sufficient flexibility for mutations that enable 
  specialized functions to evolve.
}
To inspect whether the residues comprising the parts of the network
that we predict to be crucial for the allosteric propagation and
action are significantly stronger conserved than background, we evaluate the degree of conservation of their corresponding
part of the respective protein sequence.

We first search for and retrieve amino-acid sequences that are similar
to each of the proteins considered using
Basic Local Alignment Search Tool (BLAST) \cite{Ye},
and subsequently employ
Clustal Omega \cite{Sievers} to perform a Multiple
Sequence Alignment (MSA).
In order to extract the data important for the respective protein of interest, we cut
out the part from the MSA that corresponds to the sequence of this
protein. 

We determine a background mutation rate, which is the average degree of conservation of the full sequence, i.e.\ the ratio of how often the residue in our original 
sequence occurs at the same spot in the other sequences. For each predicted source-pocket residue we divide its mutation rate by
the background mutation rate; values larger than 
one correspond to a conservation that is more than average. The
results are shown in Fig.~\ref{fig:conservation}a-c.

We find that true source-pocket residues are conserved more than
the average, with a 
few exceptions (see Fig.~\ref{fig:conservation}a). 
The beads we predict as potential source beads are conserved with 
the same, or an 
even slightly higher rate  (see Fig.~\ref{fig:conservation}b). 
This may be interpreted as that not all of 
the residues in the source 
pocket are important for allostery even though the ligand binds to them.
This idea is strengthened in context of findings in 
Appendix~\ref{subsec_pocket_size}, where we find that not all of the source beads
directly contribute to the energy input. 
When constraining combinations of smaller subsets (i.e.\ randomly chosen triplets) 
of beads within the source pocket,
we instead find that some beads do not lead to an energy change during the 
constraining step but instead
only during the relaxation step,
i.e.\ when the truly critical beads follow their displacement.
Moreover, sites that are
involved in efficient energy transmission---beads connected by a spring that
  contributes strongly to the energy change during relaxation---are as well conserved more than
the average  (see Fig.~\ref{fig:conservation}c). 
This confirms \emph{point 4} of our allosteric-lever
hypothesis. \blue{Notably, the identified efficient-energy-transmission
  beads are potentially related to the so-called ``dynamic allosteric
  residue coupling (DARC) spots'', whose mutations were hypothesized
  to cause changes in the allosteric networking \cite{Ozkan_Rev}.}
\red{This feature may also be a coincidence, as mutations in proteins
  are selected for many reasons and ENM do not explicitly encode
  any chemical specificity.}

\section{Discussion and outlook}
Allosteric mechanisms in most instances remain a biophysical enigma
that elude a general, predictive description
\cite{Motlagh}. Notwithstanding the advances in the field, so far
not much success has been achieved in identifying a set of
quantitative and transferable ``ground rules'' that would explain how
allostery works \cite{Hilser}. As already mentioned, this may be due
to the nonexistence of discrete categories of allosteric motions,
i.e.\ there is a continuum of different mechanisms available for
proteins to choose from \cite{Liu_Doing_2021}.

Meanwhile, it has long been known that proteins display certain common
features: They have 
similar vibrational spectra from low to intermediate frequencies
\cite{Avraham}, 
the associated soft modes that propagate over the entire length scale tend
to be relevant for protein function \cite{Nicolay,Na}, and
harder modes typically localize at smaller length scales near
rigidly-connected 
regions \cite{Bartlett,Yuan} 
as a result of inhomogeneities \cite{Anderson}.

We found the
allosteric response in ENMs to emerge from an intricate
coupling
between a mixture of harder
modes that transmit input displacements, via a fine-tuned combination of soft modes,
from the allosteric source
pocket to a specific rearrangement of the target pocket. 
Apparently a strong perturbation is \emph{not} required to induce the
mode mixing giving rise to allosteric
effects. \blue{Protein dynamics is thought to be evolutionarily
  optimized to enable allosteric behavior \cite{Bahar_Evo}.}
In fact, vibration spectra of proteins appear to be
poised near criticality \cite{Kaneko,Granek_ENM,Granek_criticality}.
It is thus conceivable that
small mutations give rise to structures that enable the communication of soft and
hard modes already at small external perturbations, such as ligand
binding, at specific locations.  
Thereby it
is possible to impose directionality and specificity through a balanced
coupling of stiff and soft modes. \blue{Moreover, mutations of
``dynamic allosteric
  residue coupling (DARC) spots'', which display medium flexibility but high
  coupling to active sites, were suggested to be mainly responsible for altering protein
  function \cite{Ozkan_Rev}.} This \blue{combined} in turn allows for a precise control
over protein function. 

Given that  the ``physical principle of allostery'' is
\emph{neither} a set of structural motifs \emph{nor} motion
patterns, we therefore proposed it to correspond to a common pattern in
the coupling of many complex modes, which is our \emph{allosteric-lever
hypothesis}. Accordingly, a structural 
perturbation of the allosteric source site (i.e.\ due to ligand
binding) preferentially loads localized stiff modes and relaxes
along an optimized mixture of softer modes that specifically rearrange
the target. The hypothesis was confirmed by extensive analysis of
proteins with known structures and allosteric pairs as well as by
means of pseudoproteins,  trained to display an allosteric
response.  

Our proposition stresses the fundamental difference between 
simply cooperative systems with single-mode responses
\cite{Yan_Ravasio_Brito_Wyart_Principles_2018,Ravasio_Flatt_Yan_Zamuner_Brito_Wyart_Mechanics_2019}
and specific, functional responses of proteins involving coordinated motions
\cite{Miyashita_Onuchic_Wolynes_Nonlinear_2003,Kim_Lu_Strogatz_Bassett_Conformation_2019},
as well as 
the non-linearity due to mode coupling and the non-reciprocity of the
allosteric response as a consequence of a multi-mode
response. Accordingly, allostery is, at least to some extent, a
``non-equilibrium'' phenomenon in the sense that it only fully
manifests upon perturbation.

Our results support the idea of allostery being a property
of all proteins that is merely amplified in
``allosterically active'' ones \cite{Daily_Gray_Local_2007}. 
Moreover, it is also conceivable
that mutations of amino acids alter the elastic eigenmodes of the
resting network as well as how the
perturbation of the source site couples the eigenmodes. 
This would in turn provide, alongside the so-called
``sector-connectivity picture'' \cite{Rocks_Ronellenfitsch_Liu_Nagel_Katifori_Limits_2019,Rocks_Liu_Katifori_Hidden_2021}, the elusive mechanism of the 
observed energetic connectivity pathways in proteins
\cite{pathways,pathways_2}.

However, even if the allosteric lever is internally consistent and also
seems to agree with existing observations, we may not yet call it a
``principle'', at least not in the 
mathematico-physical
sense. In order to 
attempt a (still informal) \blue{first} step in this
direction, we may proceed as 
follows. We assume for simplicity and without serious loss of generality that for given network an input perturbation of the
source $\bc(k),\,k=1,\ldots,N_{\rm step}$ yields a unique
``trajectory'' of instantaneous Hessians $\Hess_k$ and hence also a
unique 
response trajectory $\bR(k)$. For a fixed input perturbation
$\{\bc(k)\}$ 
the principle of allostery should then
correspond to a solution of the joint optimization problem that
maximizes $\{\delta U(\delta\bC_k)\}$ and minimizes $\{\delta
U(\delta\bS_k)\}$ (see
Eqs.~(\ref{takeup}-\ref{relax})) conditioned on the specific/desired response of the
target $\mathbf{T}_{*}^{\rm fin}$, i.e.\ 
\begin{equation*}
\sup_{\{\HC_k\}}\delta\bC^T_k\HC_k\delta\bC_k\,\land\!
  \inf_{\{\HC_k,\HB_k,\HS_k\}}\!\delta\bC_k^T(\HC_k-\HB_k\HS_k^{-1}\HB_k^T)\delta\bS_k\,\big
  |\,\mathbf{T}_{*}^{\rm fin}
\end{equation*}  
where $\{\HY_k\}$ denotes the entire trajectory of the sub-matrix $\HY_k=\HC_k,\HB_k,\HS_k$ for
$k=1,\ldots,N_{\rm step}$. The above optimization ``principle'' may correspond to a
mathematical definition of allostery involving a conformational
change. This idea will be tested in forthcoming publications using
training trajectories of pseudoproteins. Moreover, it would also
be interesting to incorporate reversible, dissociable bonds between
springs \cite{Poma_Li_Theodorakis_Generalization_2018} and to address
allostery without conformational change 
\cite{Cooper_Dryden_Allostery_1984}. 

\blue{On a more general note, ENMs face well known limitations when dealing with extremely large 
  conformational or structural changes cases involving the fold of the protein. 
  Employing smart modifications,
  scientists have been able overcome
  some of these limitation and employed
  ENMs to predict breaking points during
  unfolding 
  \cite{Su_Li_Hao_Chen_XinWang_Protein_2008}.
  While our computational framework can
  in principle already adapt to
  topological changes through dynamic
  connectivity reassessment, systems
  with distinct structural states may
  require integration with complementary
  methods, such as MD simulations or MS-ENMs. 
  However, these limitations primarily affect
  the detailed characterization
  of the response rather than the
  identification of allosteric sites, as
  the underlying principle of the
  allosteric lever is independent of the
  scale of the motion.
}

An exciting direction of future research would be the application of
the concepts developed here in the
computational design of
 allosteric drugs   
 \cite{Nussinov_Tsai_Allostery_2013,Chatzigoulas_Cournia_Rational_2021}.
 Unlike drugs that bind to a known target, allosteric sites are often
 unknown and the modulatory effect of the binding is \emph{a priori} difficult to predict
 \cite{Nussinov_Tsai_Allostery_2013}. Moreover, a chemically detailed
 representation of the binding
must be considered in the
evaluation of the protein's structural response, which is
computationally extremely demanding. The
method presented here may be used in a high-throughput screening that identifies
allosteric source-site candidates for a given desired response, either by constructing ENMs as here
or by determining the local Hessian from a more detailed
interaction potential. 
The selection of identified source sites can henceforth
be refined and optimized with state-of-the-art docking methods
\cite{Chatzigoulas_Cournia_Rational_2021}.

\section{Data availability}
The source code allowing to reproduce all results presented here
and some additional extensions is available 
upon request from the corresponding author 
and will become openly accessible 
at \url{https://github.com/maxvossel/elastory} upon publication of the article.
.

\subsection*{Acknowledgments}
The authors thank Dima Makarov, Gerhard Stock, David Hartich, Cai Dieball, and Lars Bock for insightful discussions.
The financial support from the German Research Foundation (DFG) through the Emmy Noether
Program (grant GO 2762/1-2) and the Heisenberg Program (grant GO 2762/4-1)
to AG is gratefully acknowledged.

\appendix
\section{Binding-pocket candidates in
  proteins}\label{Appendix-binding}
\setcounter{equation}{0}
\setcounter{figure}{0}
\setcounter{table}{0}
\renewcommand{\thefigure}{A\arabic{figure}}
\renewcommand{\theequation}{A\arabic{equation}}
\renewcommand{\thetable}{A\arabic{table}}

Once the protein-derived network has been set up, we must identify
possible binding-pocket candidates. 
Fortunately, this is an extensively investigated problem in structural
biology. 
We adopt the algorithm proposed in \cite{tian2018castp} that is based on a purely
geometric arguments and requires no chemical details to predict pockets.
It is based on 
``the rolling ball'' algorithm \cite{Shrake} frequently used to
determine the Solvent
Accessible Surface (SAS) (also called Connolly Surface
\cite{Connolly}). 
The surface is estimated with a probe of the size of the solvent which
samples the molecular surface along the respective van der Waals
radii of the atoms.  This is essentially equivalent to rolling a ball
along the surface. A Delaunay triangulation \cite{Delaunay}
(i.e.\ the most efficient way to partition
the convex hull of a set of points in 2 or 3 dimensions into triangles or tetrahedra)
using the positions of atoms is then compared 
with the solvent accessible surface \cite{Edels_1994},
and allows to define
the potential pockets using discrete flows \cite{Edels_1993},
a method to discriminate inner and outer cells of the Voronoi diagram
\cite{Liang_a,Liang_b}. The algorithm is
openly available via a web interface and returns a list of residues
in each determined binding pocket.
\begin{figure} \centering
  \includegraphics[width=.5\textwidth]{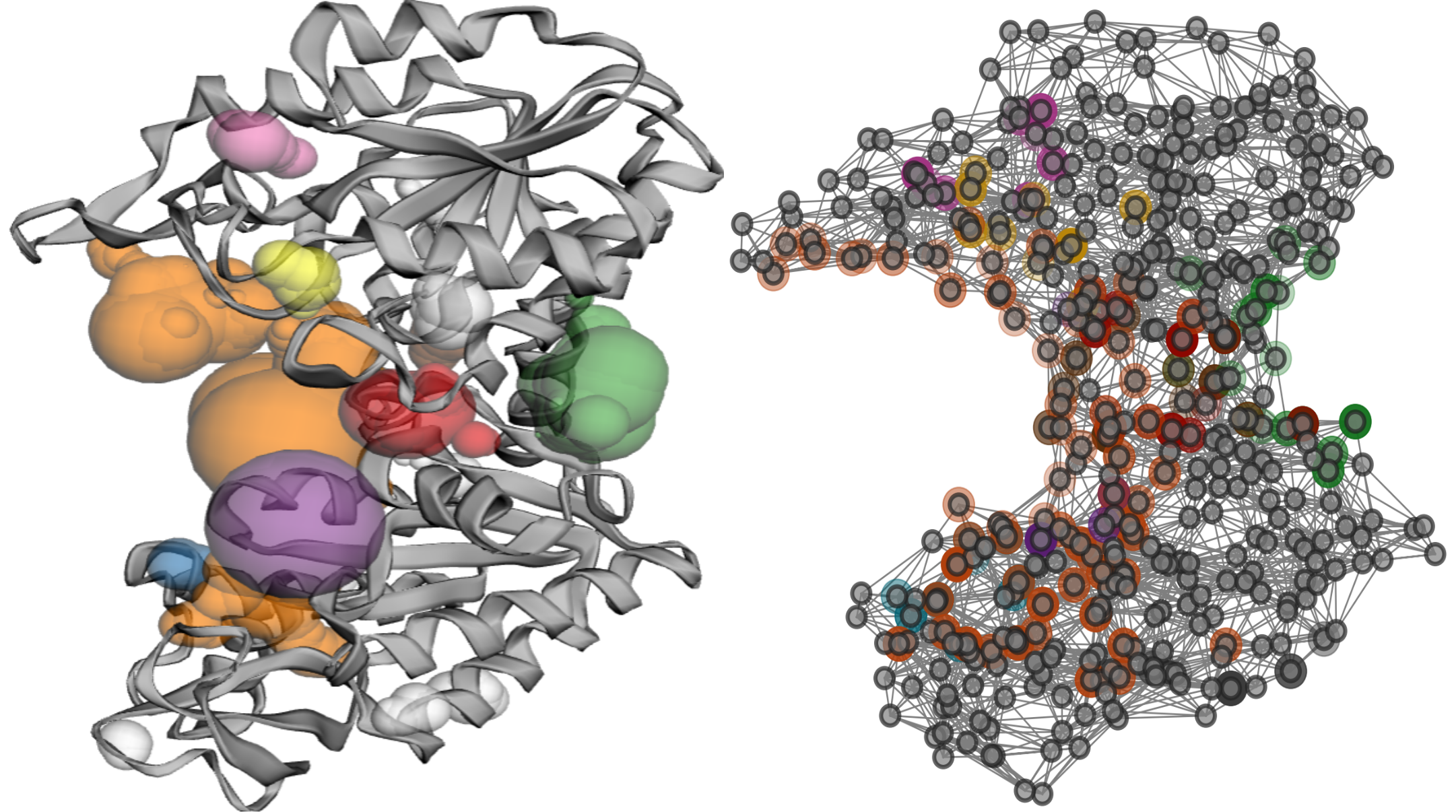}
  \caption[Pocket candidates cartoon and ENM]{
  \label{fig:pockets}
    Predicted binding-pocket candidates in the proteins are determined geometrically, based on an algorithm from~\cite{tian2018castp}.
    The 9 largest out of 77 total pockets shown.
    (Left) Cartoon representation of HSP90 (PDB ID: 2ior~\cite{shiau2006structural}) as an example protein. (Right) The beads ($\alpha$-carbons) constituting these pockets are colored accordingly in the {ENM}.
  }
\end{figure}


\begin{table}
  \centering
  \caption{Parameters of proteins analyzed in this work:
    $r_c$ is the cutoff length in units of \AA; $N$ is the total
    number of beads, $N_T$ and $N_S$ the number of beads comprising
    the target and source pocket, respectively, and $N_{\text{s}}$ and $N_{\text{p}}$ the numbers of pocket candidates, subsampled pockets and original pockets, respectively.}
  \vspace{10pt}
  \label{tab:prots}
  \begin{tabular}{llllllllll}
   \hline
  Name & PDB &state& $r_c$&$N$&$N_T$&$N_S$&$N_{\rm s}$&$N_{\rm p}$& Reference   \\ \hline\hline
  HSP90 & 2ior & holo &  9  & 228 & 3  & 9  & 1397 & 33  & \cite{shiau2006structural} \\ 
  MetJ  & 1cmc & apo  &  10 & 104 & 3  & 10 & 640  & 23  & \cite{rafferty1989three} \\
  PurR  & 1wet & holo &  11 & 338 & 27 & 5  & 2776 & 48  & \cite{schumacher1997x} \\
  HSA   & 1n5u & holo &  9  & 583 & 16 & 14 & 3523 & 51  & \cite{wardell2002atomic} \\ 
  HSA   & 2bxd & holo &  9  & 578 & 16 & 8  & 4169 & 77  & \cite{ghuman2005structural} \\ 
  HSA   & 2bxg & holo &  9  & 578 & 16 & 14 & 3168 & 68  & \cite{ghuman2005structural} \\ 
  ADK   & 4ake & apo  &  8  & 214 & 4  & 5  & 1390 & 27  & \cite{muller1996adenylate} \\
  ADK   & 1ake & holo &  8  & 214 & 4  & 5  & 1145 & 29  & \cite{muller1992structure} \\
  MyoV  & 1oe9 & apo  &  9  & 730 & 3  & 31 & 5192 & 110 & \cite{coureux2003structural} \\
  MyoV  & 1w7j & holo &  9  & 752 & 3  & 31 & 6065 & 116 & \cite{coureux2004three} \\
  PGK1  & 4o33 & holo &  9  & 417 & 6  & 21 & 3029 & 53  & \cite{chen2015terazosin} \\
  PDK1  & 3hrf & holo &  14 & 287 & 4  & 10 & 1388 & 24  & \cite{hindie2009structure} \\
  LacR  & 1efa & holo &  10 & 328 & 10 & 14 & 2154 & 47  & \cite{bell2000closer} \\
  LacR  & 1tlf & apo  &  10 & 296 & 10 & 14 & 1161 & 47  & \cite{friedman1995crystal} \\
  \hline
  \end{tabular}
  \end{table}
  
  \blue{
  A minor limitation of our approach is, that binding pocket candidates are solely based on the initial configuration of available protein structures. 
  However, there are multiple options available to overcome this
  limitation in future studies, initial work in this direction is already ongoing:
  One could use previously identified cryptic sites \cite{modi2021protein} together with their corresponding configurations of the proteins.
  One could also use Molecular Dynamics simulations to sample different conformations of the protein 
  and identify new pockets as well as new initial configurations for
  the ENM. 
  Another option would be to iteratively add new potential
  binding-site candidates during the response calculation,  
  i.e.\ repeating the search for binding sites after each step of the response calculation.
  It would be very interesting to see, e.g.\ once cryptic sites have
  been found using the above methods,  whether the same principle that
  we  found here also governs the communication towards those active sites.
}

\subsection{Ruling out pocket-size effects}\label{subsec_pocket_size}

We find that some of the larger pockets predicted by the algorithm are
huge, encompassing up to half of the proteins, see Table~\ref{tab:prots}. 
To account for the numerous possible ways a
ligand could bind within these, we draw combinations of beads out of all the beads in the
predicted pocket. These combinations consist of three beads, 
and per pocket 125 triplets are drawn randomly, leading to an increase
in the number of scanned pockets by a factor of 20-50.~For the total
number of ligand-binding combinations see Table \ref{tab:prots}.
An example of binding pockets predicted by the web server for the Human
Serum Albumin protein (HSA) \cite{ghuman2005structural} is shown as
red spheres in Figs.~\ref{fig:example_intro}a. and \ref{fig:pockets}.

\begin{figure}[!htbp]
  \centering
  \includegraphics[width=.5\textwidth]{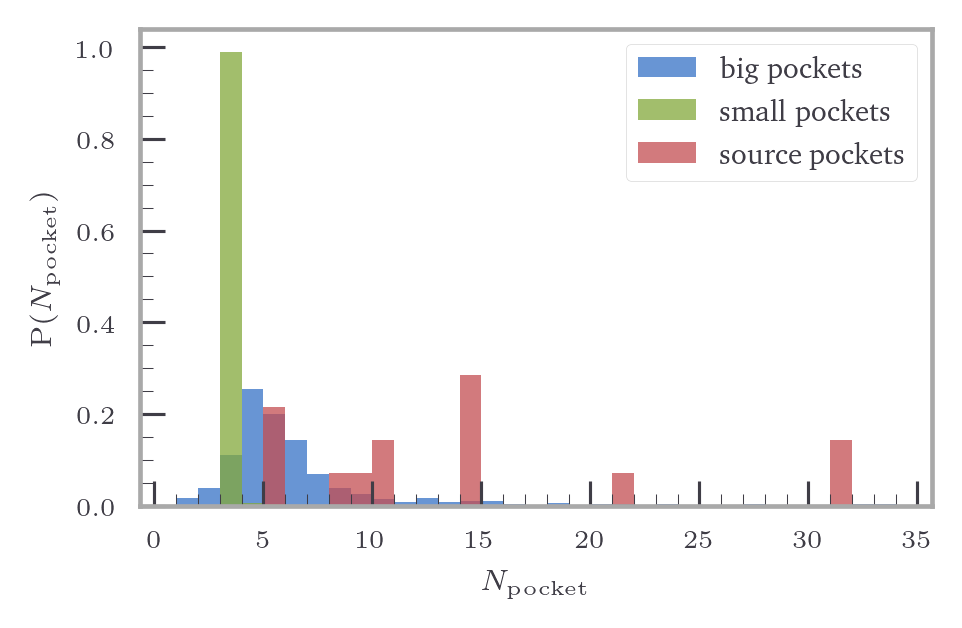}%
  \caption{
    \label{fig:proteins_pocket_sizes}
    Distribution of sizes of predicted pockets in proteins.}
\end{figure}

\begin{figure}
  \centering
  \includegraphics[width=.5\textwidth]{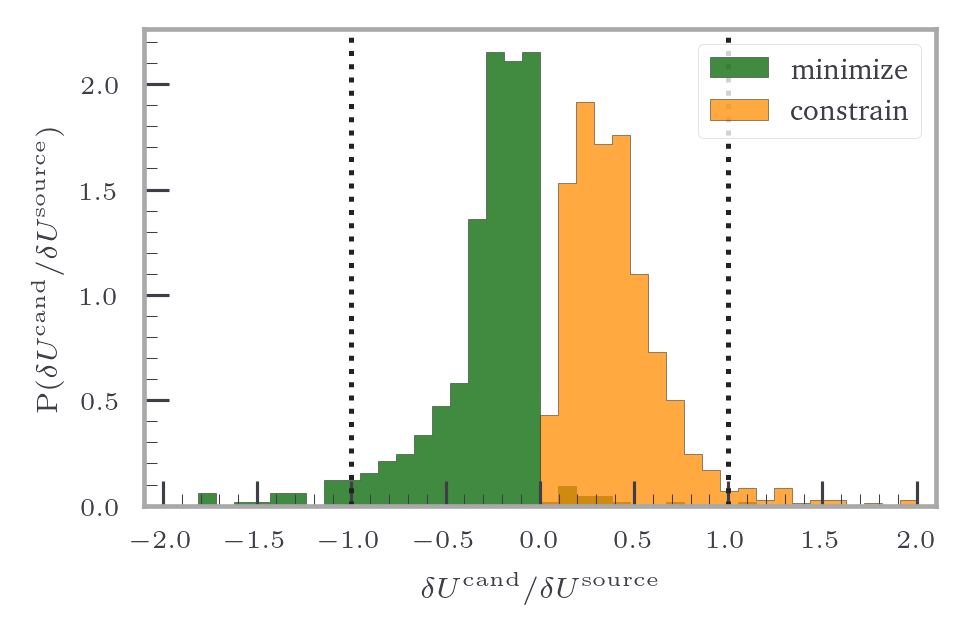}%
  \caption{
    \label{fig:proteins_energy_1d_hist_big_pockets}
    \textbf{Perturbing entire large pockets.} Histogram of energy changes during loading and relaxation for
    protein-derived networks. Here we constrain the predicted large
    (i.e.\ \emph{entire}) pockets instead of sub-sampling beads comprising them.}
\end{figure}

As mentioned in Sec.~\ref{Lever}, in order to rule out possible effects caused by
different pocket sizes, we directly perturb the large predicted
pockets and study the energy changes accompanying their perturbation/constraining.
We find \emph{no} qualitative difference between scanning smaller
parts of these binding pockets and entire pockets, as shown in Figs.~\ref{fig:proteins_energy_1d_hist_big_pockets}.
The exceptionally high values of energy uptake and release when
constraining the true source pocket in contrast to almost all other
pockets is still very well pronounced, thus excluding a possibly trivial pocket-size effect.
To eliminate any further concerns regarding pocket-size effects, we
take a closer look at the true source pocket itself. Indeed, we did
not scan in Sec.~\ref{Appendix-binding} large entire pockets, 
but instead generated various combinations of possible ligand-binding
sites within these.
The true source pocket is somewhat larger than these considered
combinations, see Fig.\,\ref{fig:proteins_pocket_sizes}. As we
throughout always divided by the number of perturbed beads, their
actual number (i.e.\ size) should in principle have no effect. 

However, a further control is advisable to rule out other
possible pocket-size effects occur. Equivalently to the sampling
performed within the larger (predicted) binding-pocket candidates, we
therefore also inspect the source 
pocket by scanning across \emph{all} combinations of bead triplets within the pocket.
\begin{figure}
  \centering
    \includegraphics[width=.5\textwidth]{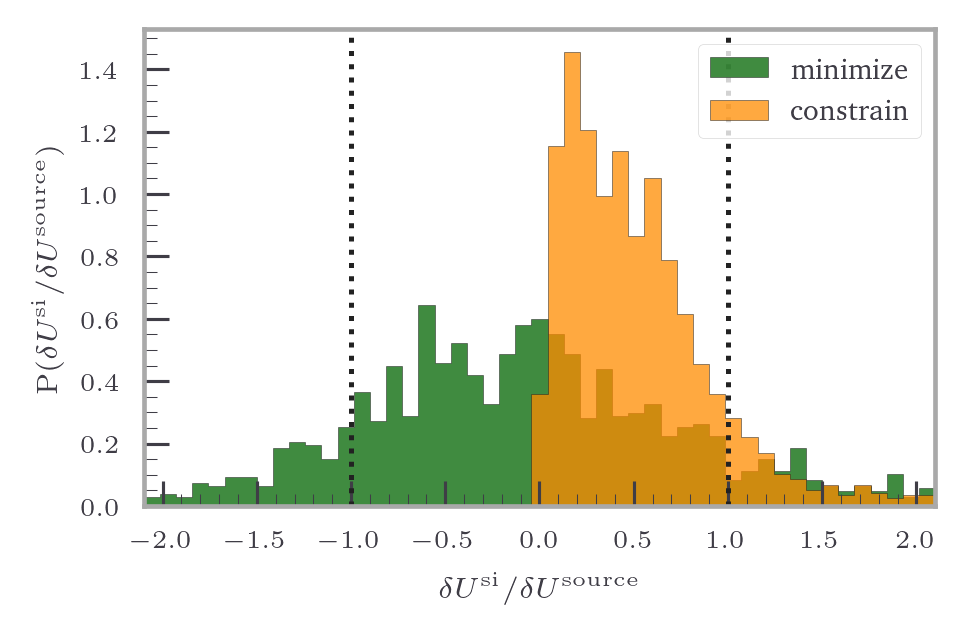}%
  \caption{
  \label{fig:proteins_energy_1d_hist_subsample_source}
  \textbf{Perturbing only subsets of large pockets.} Histogram of
  energy changes for protein-derived networks, where we sub-sample
  large source pocket candidates.}
\end{figure}

In general the fact that beads in a source pocket bind the ligand does \emph{not} imply that all of them are equally important for transmitting the allosteric signal.
It is just as conceivable that only a subset thereof is actually
responsible for loading the stiff collective springs. 
This is also how we interpret the results of sub-sampling the true source
pockets shown in
Fig.~\ref{fig:proteins_energy_1d_hist_subsample_source}, where
we see a significant difference in the distribution of energy changes
during the relaxation step of the true
with respect to the other pocket candidates. 
Notably, however, we recover the familiar trend for the energy change during loading.
This implies that indeed not all beads in the true source pocket
contribute equally/comparably to the loading.
Rather there exists a subset of critically important beads within the source pockets, which account for the major energy input during loading.
We observe that as long as we do not constrain these critical beads and
instead the beads connected to these critical ones, 
the energy does not increase immediately during the loading step, but
instead in the following minimization steps, that is, when the
critical beads follow. 
In this respect, the relaxation movements of critical beads
compensate for the absence of a large/efficient loading of
non-critical beads in the source pocket.

\section{Construction and training of artificial allosteric pseudoproteins}\label{Appendix-training}
\setcounter{equation}{0}
\setcounter{figure}{0}
\setcounter{table}{0}
\renewcommand{\thefigure}{B\arabic{figure}}
\renewcommand{\theequation}{B\arabic{equation}}
\renewcommand{\thetable}{B\arabic{table}}

Many alternative methods are available for designing and training
allosteric networks. In our work we generalize and optimize the method developed by
Flechsig \cite{Flechsig_Design_2017}. We first ``grow'' the networks to resemble folded coarse-grained
proteins. The first bead is positioned at the origin and afterwards
the following steps are iterated until the desired total number of
beads $N$ is reached: Each subsequent
bead is placed on the surface of a $3-$dimensional sphere centered at
the location of the previously positioned bead with a radius $d_{\rm
  min}$ under the constraint
that the distance to all other beads is also above $d_{\rm min}$.
 Two other constraints are imposed: First, all beads must lie
within a large sphere of radius $d_{\rm big}$ such that
a globular shape is attained. Second, the volume enclosed by two
smaller spheres of $d_{\rm small}$ which intersect the large sphere is
left unpopulated, such that two ``binding pockets'' emerge. 
For a subset of networks we combine two of the grown structures
by overlapping them 
and removing all beads that thereafter lie closer together than $d_{\rm min}$. 
Detailed parameters for the grown
networks are given
in Table~\ref{tab:nets}. 
The source and
target pocket, respectively, are
comprised of two or three
beads which are selected by hand.

These grown networks are trained to exhibit a long-range allosteric
effect consisting of an 
input and an output. A ligand-binding event that closes the source pocket is mimicked by
pulling the beads of the source pocket towards their local center of
mass---we refer
to this as \emph{the input}. Conversely, we refer to the opening or closing
of the target binding pocket (depending on the variant being trained) 
as \emph{the output}. The full nonlinear response of the network to
the input perturbation of the
source pocket is described below.

We quantify the  state of the source ($A=S$) and target ($A=T$)  pocket, respectively,
via the squared radius of gyration, defined as
\begin{equation}
r^2_{\rm gyr}(A)\equiv \frac{1}{N_A}\sum_{i\in A}\Big(\br_i-\sum_{j\in A}\br_j/N_A\Big)^2
  \label{rgyr}
\end{equation}
We train the networks according to their response to the input
perturbation in an open-system Monte Carlo
(MC) fashion. That 
is, beads can change their resting position $\br_i^0$ and new beads
can be created or existing ones removed. The algorithm is inspired by
the principle of natural evolution, corresponding to
mutations that change amino acids at a given position in the sequence, or more drastic
insertions \cite{King} and deletions \cite{Kimura}, respectively. 

First, one of the three possible steps (move, delete, insert) is randomly
chosen with probabilities $1/2,1/4$ and $1/4$, respectively, and the
response of the mutated network determined. 
If the response at the target site has improved according to the
variant trained (i.e.\ $r^2_{\rm gyr}(T)$ has increased or decreased
for a symmetric and antisymmetric response, respectively), the change
is accepted with probability one and otherwise with probability
$\exp(-|\Delta r_{\rm gyr}(T)|)$, where $\Delta r_{\rm gyr}(T)\equiv
r^{\rm init}_{\rm gyr}(T)-r^{\rm fin}_{\rm gyr}(T)$ is the
change in target-pocket size during the response. The
exponent prevents too large unwanted changed while concurrently
allowing to avoid getting stuck in local favorable but
sub-optimal minima. Less than $10^3$ MC steps were required to
incorporate the desired effect 
into the networks.

Notably, to remove any potential bias in the above method, we also considered
networks derived from the random “dense packed-spheres” algorithm
\cite{Baranau}. This yielded structures that are closer to those in \cite{Yan_Ravasio_Brito_Wyart_Architecture_2017,Rocks_Pashine_Bischofberger_Goodrich_Liu_Nagel_Design_2017}.
However, we subsequently subjected these structures to the same
evolutionary training, since simple flipping or pruning of bonds is
not possible within the standard ENM framework following Tirion
\cite{Tirion_Large_1996}.  
In total, we generated a set of 30 artificial ENMs
trained to display a specific response.

\section{Binding-pocket candidates in artificial networks}\label{Appendix-pockets}
\setcounter{equation}{0}
\setcounter{figure}{0}
\setcounter{table}{0}
\renewcommand{\thefigure}{C\arabic{figure}}
\renewcommand{\theequation}{C\arabic{equation}}
\renewcommand{\thetable}{C\arabic{table}}
In the case of artificial networks we identify binding-pocket
candidates as pairs or triplets of beads resembling a  simple binding
pocket. We impose (i) that binding-pocket candidates must be accessible to
hypothetical ligands (i.e.\ that they must lie on the surface of the
network) and (ii) that they actually form a pocket, that is, that is the lines connecting
the respective beads do not cut through 
the bulk of the network.  
Both constraints embody the definition of the surface of a
network. As opposed to atomically detailed proteins above, defining the surface of a network (i.e.\ the concave hull of a cloud of points in 3
dimensions) is neither mathematically nor intuitively well posed. 
Many different shapes may envelop the respective structures.

\begin{figure} \centering
    \includegraphics[width=.5\textwidth]{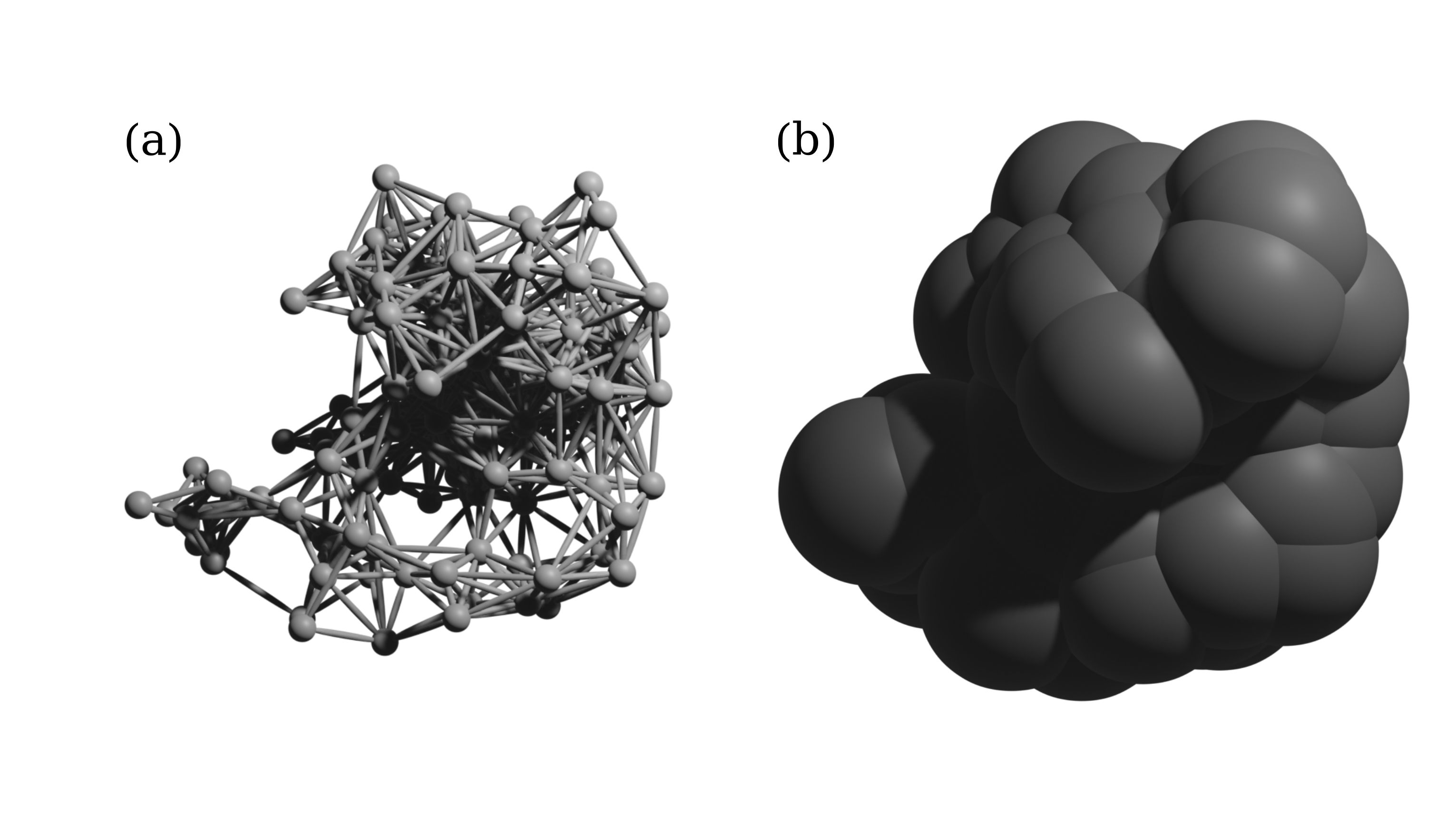}
    \caption[Network surface]{\label{fig:net_to_surf}
    We define the surface beads of a network (a) as those beads that contribute to the shared surface after growing the radii of the beads (b). Potential binding pockets are selected from this set of beads.}
  \end{figure}

We chose a unique definition of ``the surface'' as the surface that emerges
by growing a sphere around each of the beads such that a coherent solid
body emerges. This fused-spheres object consistently defines a surface
of the network that actually is quite similar to the definition of the SAS area for
proteins (see Fig.\,\ref{fig:net_to_surf}). We declare beads 
that belong to spheres participating in this outer shared surface as surface beads and the
remaining ones as interior beads.  

Pocket pairs and triplets are now selected from the set of surface beads. The condition
that binding pockets must lie on the outer surface rules out
``implausible ligand'' effects, enforced as follows. Out of all 
possible combinations of pars of surface beads and triplets we only
consider those whose direct connecting lines do not cut any of the
interior beads. In addition, we omit pairs if they are
direct neighbors or if their distance is above a certain threshold
value.

\begin{table}[h!]
  \centering
  \caption{Overview of parameters of analyzed pseudoproteins; $r_c$ is
    the cutoff length (in natural, dimensionless units), $N$ the total
    number of beads, $N_T$ and $N_S$ the number of beads composing the
    target and source pocket, respectively, and $N_{\text{sp}}$ and $N_{\text{st}}$ the numbers of pocket candidates, bead pairs and triplets, respectively.}
  \label{tab:nets}
  \begin{tabular}{lllllllll}
    \hline
  ID & variant & $r_c$ & $N$ & $N_T$ & $N_S$ & $N_{\rm sp}$ & $N_{\rm st}$ & origin \\ \hline\hline
   1 & anti    & 1.63  & 120 &  2 & 2  & 206 &  206  & packed spheres \\
   2 & symm    & 6.0   & 95  &  3 & 3  & 376 &  376  & pseudoprotein\\
   3 & anti    & 6.0   & 94  &  3 & 3  & 295 &  295  & pseudoprotein\\
   4 & symm    & 6.0   & 89  &  3 & 3  & 224 &  224  & pseudoprotein\\
   5 & symm    & 6.0   & 142 &  3 & 3  & 201 &  201  & pseudoprotein\\
   6 & anti    & 6.0   & 142 &  3 & 3  & 309 &  309  & pseudoprotein\\
   7 & anti    & 6.0   & 180 &  3 & 3  & 530 &  530  & pseudoprotein\\
   8 & symm    & 6.0   & 93  &  3 & 3  & 461 &  461  & pseudoprotein\\
   9 & anti    & 6.0   & 154 &  3 & 3  & 315 &  315  & pseudoprotein\\
  10 & symm    & 6.0   & 100 &  3 & 3  & 540 &  540  & pseudoprotein\\
  11 & symm    & 6.0   & 99  &  3 & 3  & 401 &  401  & pseudoprotein\\
  12 & anti    & 6.0   & 94  &  3 & 3  & 294 &  294  & pseudoprotein\\
  13 & symm    & 6.0   & 93  &  3 & 3  & 535 &  535  & pseudoprotein\\
  14 & symm    & 6.0   & 147 &  3 & 2  & 495 &  495  & pseudoprotein\\
  15 & symm    & 6.0   & 94  &  3 & 3  & 330 &  330  & pseudoprotein\\
  16 & symm    & 6.0   & 99  &  3 & 3  & 332 &  332  & pseudoprotein\\
  17 & symm    & 6.0   & 154 &  3 & 3  & 465 &  465  & pseudoprotein\\
  18 & anti    & 6.0   & 89  &  3 & 3  & 288 &  288  & pseudoprotein\\
  19 & anti    & 6.0   & 93  &  3 & 3  & 461 &  461  & pseudoprotein\\
  20 & anti    & 6.0   & 95  &  3 & 3  & 317 &  317  & pseudoprotein\\
  21 & anti    & 6.0   & 99  &  3 & 3  & 400 &  400  & pseudoprotein\\
  22 & anti    & 6.0   & 100 &  3 & 3  & 480 &  480  & pseudoprotein\\
  23 & symm    & 6.0   & 135 &  3 & 3  & 571 &  571  & pseudoprotein\\
  24 & symm    & 6.0   & 88  &  3 & 3  & 187 &  187  & pseudoprotein\\
  25 & symm    & 6.0   & 87  &  3 & 3  & 244 &  244  & pseudoprotein\\
  26 & symm    & 6.0   & 169 &  3 & 3  & 386 &  386  & pseudoprotein\\
  27 & anti    & 6.0   & 147 &  3 & 2  & 427 &  427  & pseudoprotein\\
  28 & anti    & 6.0   & 99  &  3 & 3  & 353 &  353  & pseudoprotein\\
  29 & anti    & 6.0   & 87  &  3 & 3  & 234 &  234  & pseudoprotein\\
  30 & anti    & 6.0   & 169 &  3 & 3  & 382 &  382  & pseudoprotein\\
  1F & anti    & 9.0   & 200 &  2 & 2  & 845 &  2204 & \cite{Flechsig_Design_2017} \\
  2F & symm    & 9.0   & 200 &  2 & 2  & 892 &  2691 & \cite{Flechsig_Design_2017} \\
  \hline
  \end{tabular}
\end{table}

\section{Removing rigid-body motions}\label{rigid}
\setcounter{equation}{0}
\setcounter{figure}{0}
\setcounter{table}{0}
\renewcommand{\thefigure}{D\arabic{figure}}
\renewcommand{\theequation}{D\arabic{equation}}
\renewcommand{\thetable}{D\arabic{table}}
The Hessian is singular, its nullspace $\mathcal{N}(\Hess)$ is spanned by the
vectors describing infinitesimal rigid body motions. Depending on the
number of constraints, the solution of Eq.~\eqref{soln} is not
necessarily unique, i.e.\ the system is underdetermined. The following simple
algebraic trick solves this problem.

Consider the square matrix $\HM\in\mathbbm{R}^{m\times m}$ which does not have full
rank $r$, its nullity being $n$, such that $n+r=m$.
The nullspace is $\mathcal{N}(\HM)$ with dimension $n$ and its row space (or
range) is $\mathcal{R}(\HM^T)$ with dimension $r$. 
All vectors in $\mathcal{N}(\HM)$ are
orthogonal to all vectors in $\mathcal{R}(\HM^T)$ 
Therefore, if we add a matrix given by
an outer product of a basis vector $\bx$ from the nullspace to $\HM$,
i.e.\ $\tilde{\HM}\equiv\HM+\bx\otimes\bx^T$, we increase the rank of
$\HM$ by one. Applying this for each basis vector of
$\mathcal{N}(\HM)$\footnote{Standard eigendecomposition routines are not ideal for a fast calculation of the eigenvectors that correspond to the 6 zero eigenvalues of $\Hess$. Using a shift and invert
spectral transformation from \textsf{ARPACK} \cite{Lehoucq1998ARPACKUG}, where we choose the number of eigenvalues we are interested in and provide a starting point for the search, is considerably faster. The
corresponding implementation in \textsf{scipy} \cite{scipy}
does not always converge for eigenvalues that are of the same
magnitude as the numerical precision (i.e.\ zero eigenvalues). We overcome
this issue by adding a small noise to the matrix whose nullspace is to
be determined.} 
gives $\HM$ full rank \cite{null}. This, however, does \emph{not}
change any result where the matrix is applied to a vector $\by$ from the row
space $\mathcal{R}(\HM^T)$, since 
\begin{eqnarray}
  \tilde{\HM}\by&\equiv&\HM\by+\bx\otimes\bx^T\by\nonumber\\
  &=&\HM\by+(\bx^T\by)\by=\HM\bx,\quad \forall \bx\in\mathcal{N}(\HM).
\label{add}  
\end{eqnarray}  
We apply this rank extension only
to the sub-block $\HS$ containing free beads $\bS$ in Eq.~\eqref{split}. Constraining the positions of
beads in $\bC$ would otherwise conflict the constrained degrees of
freedom. The resulting matrix has full rank, rendering the linear
system~\eqref{soln} uniquely solvable. The solution thus by
construction does not contain any rigid body motions.  

It is even possible to identify the basis vectors of the nullspace
explicitly in terms of Cartesian basis vectors $\be_x,\be_y,\be_z$. In the
3-dimensional Cartesian basis the vectors generating translations of a body with $N$
beads can be written as $3N$-dimensional supervectors,
$\btau(\alpha)=(\be_\alpha,\ldots,\be_\alpha)^T$, for
$\alpha=x,y,z$. Given the position of beads
$\bR=(\br_1,\ldots,\br_N)^T$ the basis vectors $\brho(\alpha)$ generating
infinitesimal rotations of the network can be constructed as
$\brho(\alpha)=(\be_\alpha\times\br_1,\ldots,\be_\alpha\times\br_N)^T$. Both
methods are equivalent whereby the second method is
faster. If we constrain more than two beads (essentially always true for the binding sites
in proteins) the removal of rigid body motions can be omitted entirely, speeding up the
algorithm tremendously.

\section{ROC curves for energy change during constraining}\label{app:ROC_constrain}
\setcounter{equation}{0}
\setcounter{figure}{0}
\setcounter{table}{0}
\renewcommand{\thefigure}{E\arabic{figure}}
\renewcommand{\theequation}{E\arabic{equation}}
\renewcommand{\thetable}{E\arabic{table}}
\begin{figure*}[!htbp!]
\includegraphics[width=1.\textwidth]{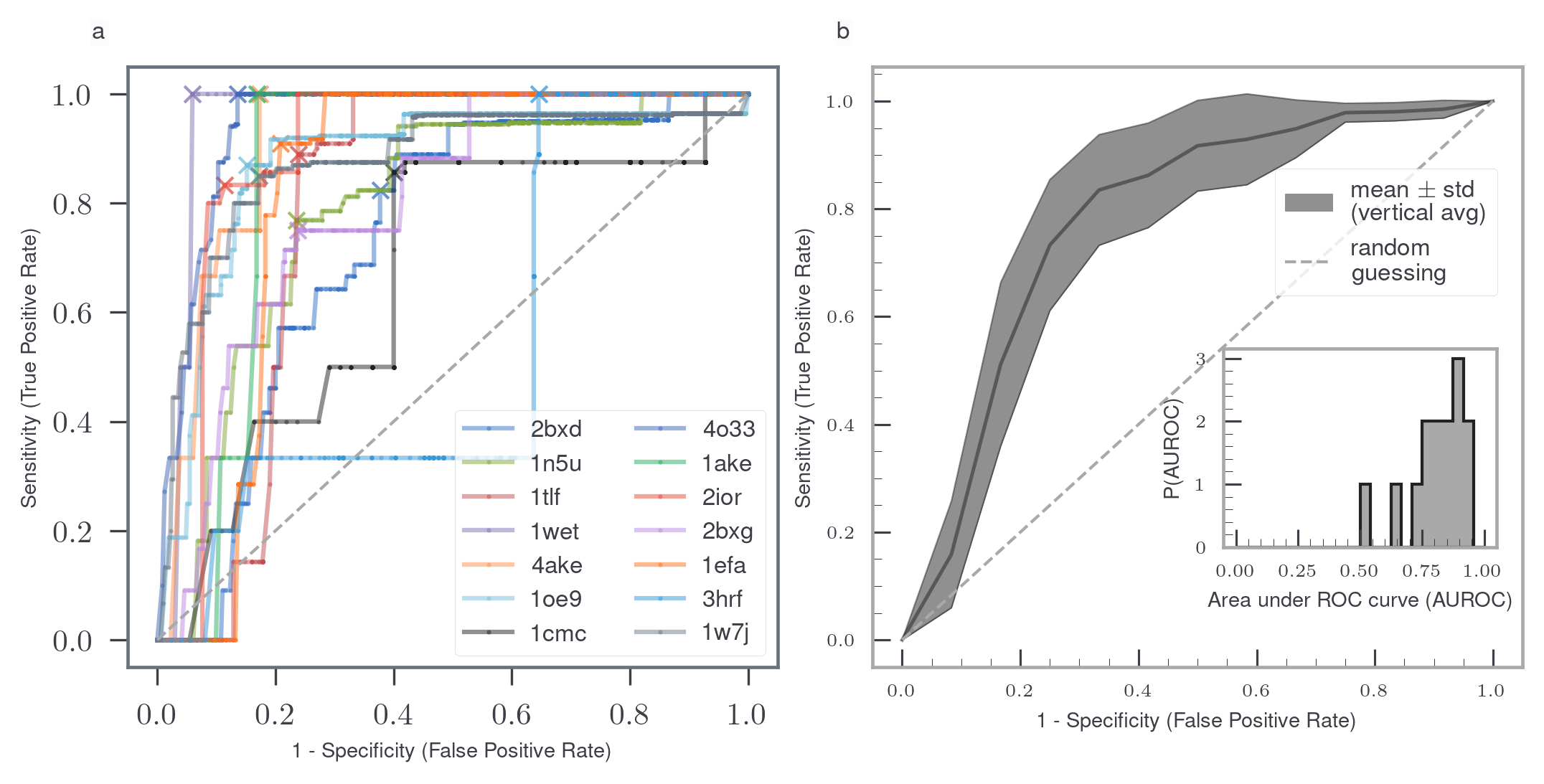}%
\caption[Predictive power - ROC curves]{
  \label{fig:ROC_constrain}
  For an explanation refer to Fig.\,\ref{fig_predict}. Results are similar on average but with a larger spread.}
\end{figure*}
For completeness, we also rank source-pocket candidates according to
how much energy is 
taken up during the constraining  step $\delta
U(\delta\bC)$. As shown in  Fig~\ref{fig:ROC_constrain},  the results are
similar on average, but show a larger variance.
\section{Finding avoided crossings}\label{Appendix_avoided}
\setcounter{equation}{0}
\setcounter{figure}{0}
\setcounter{table}{0}
\renewcommand{\thefigure}{F\arabic{figure}}
\renewcommand{\theequation}{F\arabic{equation}}
\renewcommand{\thetable}{F\arabic{table}}

\begin{figure*}[!htbp]
\includegraphics[width=1.0\textwidth]{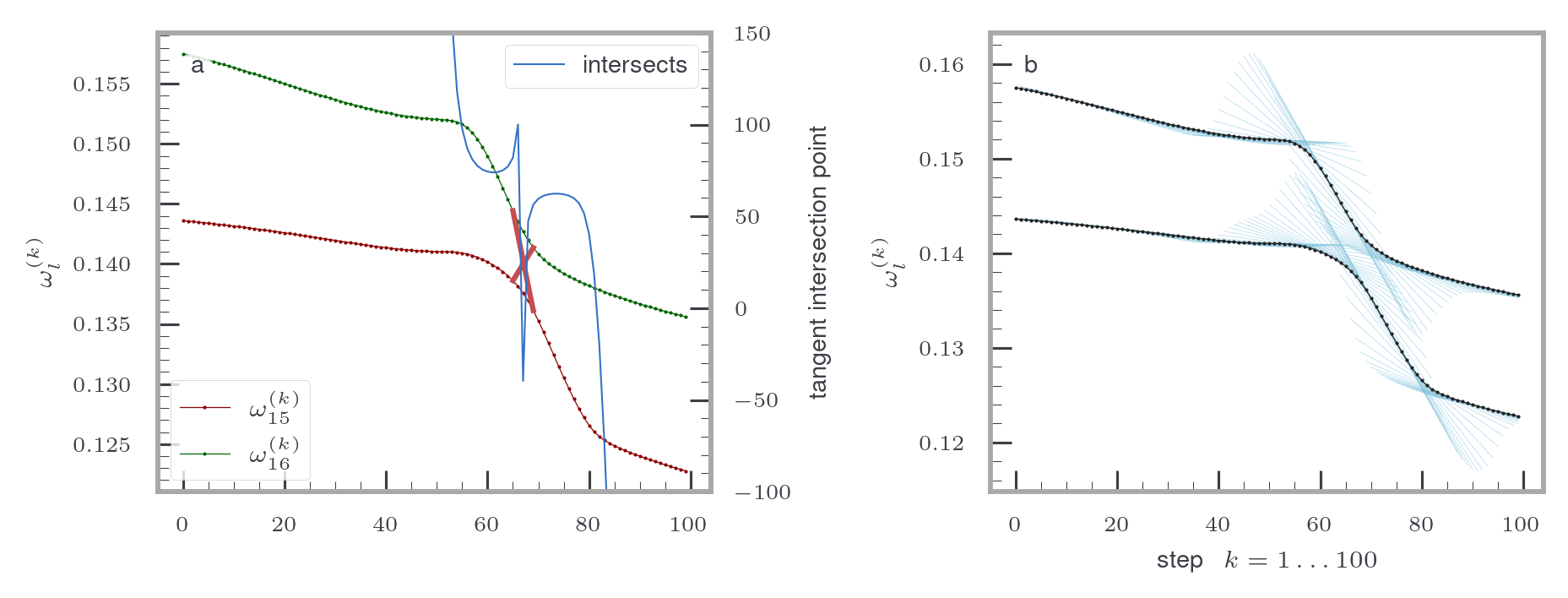}%
\caption{\textbf{Intersection points of consecutive tangents to
    eigenvalues.} An example illustrating the motion of mid-point
  tangents in Eq.~\eqref{tagent} for a set of eigenvalues during the response of one generic pseudoprotein.}
\label{avoided}
\end{figure*}
As detailed in Sec.~\ref{avoid}, we determine
mid-point tangents $T_l(k)$ for all points $k$ along the eigenvalue
curves during the response  (see Eq.~\eqref{tagent}), whereby we
exclude the first and last point of each curve. An avoided
crossing between a given pair of eigenvalues $l$
and $l+1$ at step $k$ occurs when the intersection point
first diverges towards $+\infty$ as a function of $k$ and subsequently
re-enters from $-\infty$. This is illustrated in Fig.~\ref{avoided}.
The ``precise''
location of an intersection is determined on the basis of a
sign-change of $\omega_{l+1}(k)-\omega_{l}(k)$. Only instances where the
crossing occurs first from above ($+\to -$) and then from below ($-\to +$)
are considered.

\setcounter{equation}{0}
\setcounter{figure}{0}
\setcounter{table}{0}
\renewcommand{\thefigure}{G\arabic{figure}}
\renewcommand{\theequation}{G\arabic{equation}}
\renewcommand{\thetable}{G\arabic{table}}
\blue{
\section{Correlation between non-linear response and avoided
  crossings}\label{Appendix_crossings_end}
As described in Sec.~\ref{avoid}, we also quantify the non-linearity of the response by 
the rotation of the eigenvectors at the end of the response relative to the initial state.
The average rotation of the eigenvectors at the end of the response is defined as}
\blue{
\begin{equation}
  \langle \beta^{N_{\rm step}} \rangle = \frac{1}{3N-6}\sum_{l=6}^{3N}\beta^{(N_{\rm step})}_l .
  \label{beta_end}  
\end{equation}
}
\blue{The corresponding results for $\langle \beta \rangle^{(N_{\rm step})}$
is shown in Fig.~\ref{fig:crossings_appendix}.
There is no visible difference between $\langle\beta\rangle$ and $
\langle \beta \rangle^{(N_{\rm step})}$, so non-linearity persists
throughout the trajectory.}

\blue{Careful inspection of the avoided crossings simultaneously with the occurrence of the eigenvector rotations reveals that the 
crossings \emph{always} occur together with a rotation of the eigenvectors, whereas 
the converse is not true. (The data can be made available upon request.)
Conversely, the presence of rotations without crossings indicates
complex underlying dynamics that  
cannot be explained by a simple one-to-one correspondence between the two phenomena.}

\blue{In Figs.~\ref{fig:crossings}d and ~\ref{fig:crossings_appendix}b two
outliers with a much higher non-linearity are visible, 
with PDB IDs 1oe9 and 1w7j. Both are structues of the same protein
(Myosin V motor) in different states. In the case of the two outliers,
the exceptionally high degree of nonlinearity  
possibly relates to the fundamental nature of Myosin V as a motor protein. 
Motor proteins are specifically designed to undergo large conformational changes 
as part of their function in cellular transport. 
The mechanical nature of these proteins suggests they may have evolved 
structural features optimized for significant conformational changes
during their working cycle.  
This enhanced mechanical flexibility could explain why these
structures exhibit significantly higher eigenvector rotations while
maintaining a relatively normal number of avoided crossings, setting
them apart in Figs.~\ref{fig:crossings}d and~\ref{fig:crossings_appendix}b (Fig.~\ref{fig:crossings_appendix}b also includes a legend with the PDB IDs of the proteins, colored individually).
}

\begin{figure*}[t!]
  \includegraphics[width=0.8\textwidth]{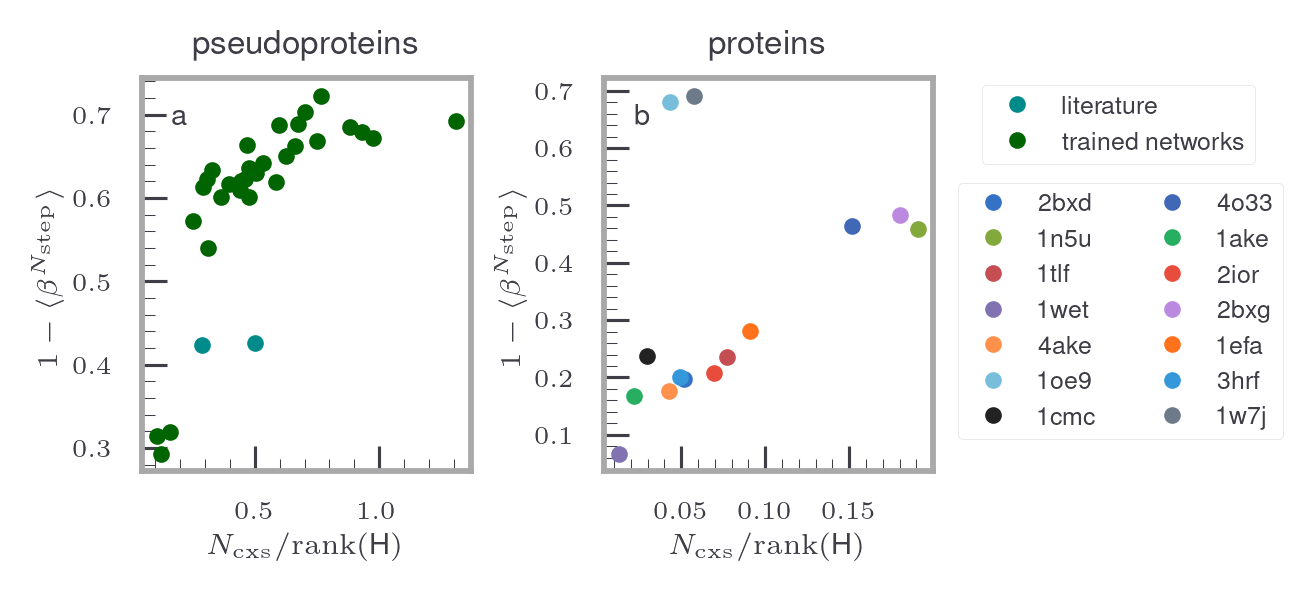}%
     \caption{\blue{\textbf{Correlation between nonlinear response and
         avoided crossings.}
      \textbf{(a-b)} Scatter plots as in
      Fig.~\ref{fig:crossings}\textbf{(c-d)} but with
      $\langle\beta\rangle^{(N_{\rm step})}$ instead of
      $\langle\beta\rangle$, separately for pseudoproteins (self-trained: green, literature: petrol) 
      and protein-derived networks (individually colored with corresponding PDB IDs in the legend).}
  }
  \label{fig:crossings_appendix}
\end{figure*}

\setcounter{equation}{0}
\setcounter{figure}{0}
\setcounter{table}{0}
\renewcommand{\thefigure}{H\arabic{figure}}
\renewcommand{\theequation}{H\arabic{equation}}
\renewcommand{\thetable}{H\arabic{table}}
\blue{
\section{Reversibility of the perturbation induced response}\label{Appendix_reversibility}
The full response at infinite resolution (for genuinely infinitesimal steps) is reversible, which may be shown as follows. Since the solution of the linear optimization problem in Eq.~\eqref{response} is step-wise unique, the trajectory obtained at finite resolution by mirroring according to  
\begin{equation}
\mathbf{R}(N-k)=\inf_\mathbf{R}\, [\mathbf{R}^T\Hess_{N-k}\mathbf{R}\,\big |\,\mathbf{R}_{i\in S}=\bC(N-k)],
  \label{rv_response}
\end{equation}
for $k\in [1,N_{\rm step}]$ is manifestly symmetric (by construction). Assuming that the continuum limit of the constraint discretization (i.e.\ the extension of $\bC(k)$ to $\bC(x)$ for real $x$) exists and that the corresponding continuous extension of the local Hessian $\Hess_{k}\to \Hess_{x}$ is element-wise continuously differentiable, then $|(\Hess_{x+{\rm d}x})_{ij}-(\Hess_{x})_{ij}|=\mathcal{O}({\rm d}x)$ and hence in the continuous limit the response in Eq.~\eqref{response} is reversible. In practice, however, one must consider discretized responses and hence we must gauge finite-resolution effects.\\      
\indent The reversibility of the iterative quadratic constrained optimization
at finite resolution  is tested by mirroring the input at the end of the response.
An example of a trajectory where the input has been mirrored at the end is shown in Fig.~\ref{fig:crossings_appendix}c.
The response is then calculated with the mirrored input, and the distance to the initial configuration is measured at the last step of the response.
The distance is measured throughout the response,
\begin{equation}
  \mathbf{D}^{(k,0)} = \frac{L_2(\mathbf{R}^{(k)} - \mathbf{R}^{(0)})
  }{ \max_{k\in[1,N_{\rm step}]}(L_2(\mathbf{R}^{(k)} - \mathbf{R}^{(0)}))},
 \label{distance_f} 
\end{equation}
and the statistics are shown in Fig.~\ref{fig:fwd_bwd_rev}a and \ref{fig:fwd_bwd_rev}b, for the pseudoproteins and protein-derived networks, respectively.
}
\begin{figure*}[ht!]
  \includegraphics[width=1.\textwidth]{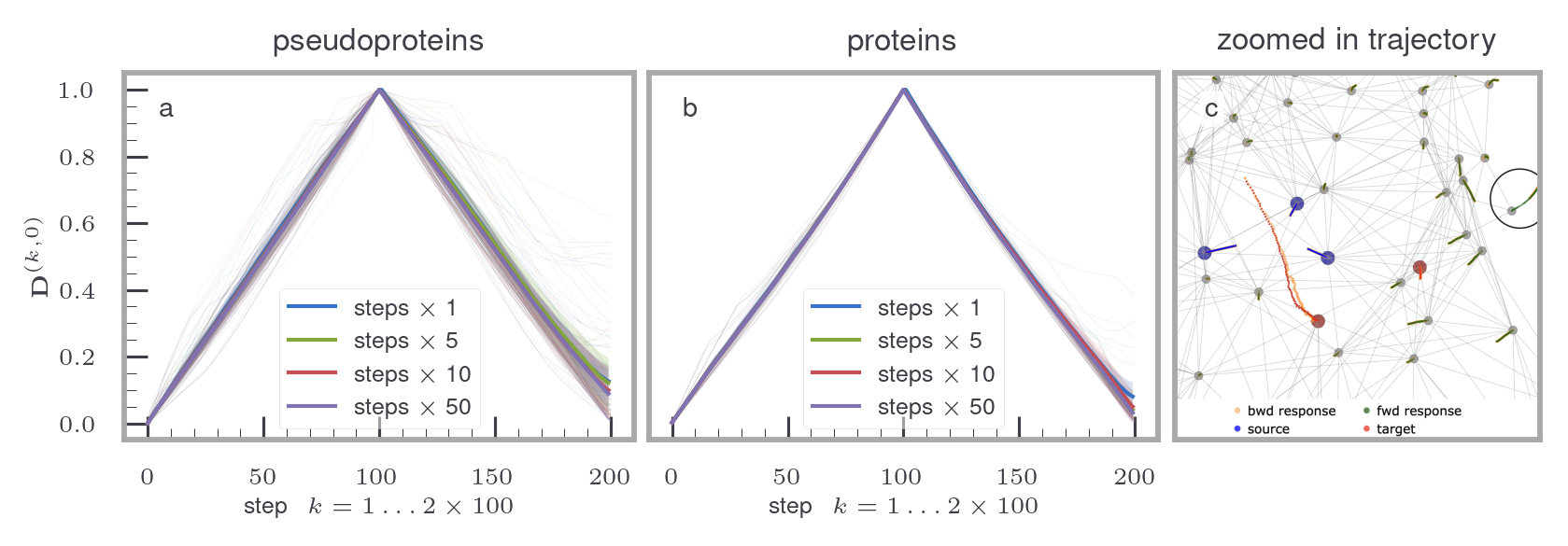}%
    \caption{\blue{\textbf{Reversibility of constrained optimization response at finite resolution.}\label{fig:fwd_bwd_rev}
    \textbf{(a-b)} Statistics of the distance $ \mathbf{D}^{(k,0)}$ in
    Eq.~\eqref{distance_f} (with mirrored input and numerically evaluated response).
    It is clearly observed that most of the structures return into a
    close vicinity of the initial configuration despite a finite step
    site and hence a discrepancy between $\Hess_{k}$ entering the
    forward (Eq.~\eqref{response}) and $\Hess_{k+1}$ entering the
    backward (Eq.~\eqref{rv_response}) response. 
    \textbf{(c)} Examplary trajectory where the input has been
    mirrored at the end. It is clearly shown that the entire network
    configuration returns into the vicinity of the initial
    configuration, within numerical errors, thereby confirming the
    reversibility of the iterative quadratic constrained optimization
    also in practice (i.e.\ at finite resolution).
    The circle indicates a bead which does not return to its initial position. 
    }
  }
\end{figure*}

\setcounter{equation}{0}
\setcounter{figure}{0}
\setcounter{table}{0}
\renewcommand{\thefigure}{I\arabic{figure}}
\renewcommand{\theequation}{I\arabic{equation}}
\renewcommand{\thetable}{I\arabic{table}}
\blue{
  \section{The effect of noise on the perturbation induced response}
  \label{Appendix_noise}
  To assess the robustness of our method against perturbations, we extended the algorithm to include the possibility of adding noise to the input perturbation. The noise is implemented as normal-distributed fluctuations of configurable strength around the original input trajectory. 
  Our investigations of ensembles of stochastic trajectories demonstrate that the method remains robust within reasonable noise levels. The energy landscape containing the response trajectory exhibits sufficient stability, characterized by a deep energy valley that effectively constrains the perturbed trajectories. As a result, even in the presence of noise, the system's response consistently follows the local minimum-energy path, quasi-statically adapting to input changes at the source site.
  This robustness suggests that our method captures fundamental mechanical properties of the system rather than artifacts of precise trajectory choice. An example of such an ensemble of noisy trajectories is shown in Fig.~\ref{fig:noise_appendix}, where multiple realizations of the response under the influence of noise demonstrate the stability of the overall response pattern.
}

\begin{figure*}[ht!]
  \includegraphics[width=0.5\textwidth]{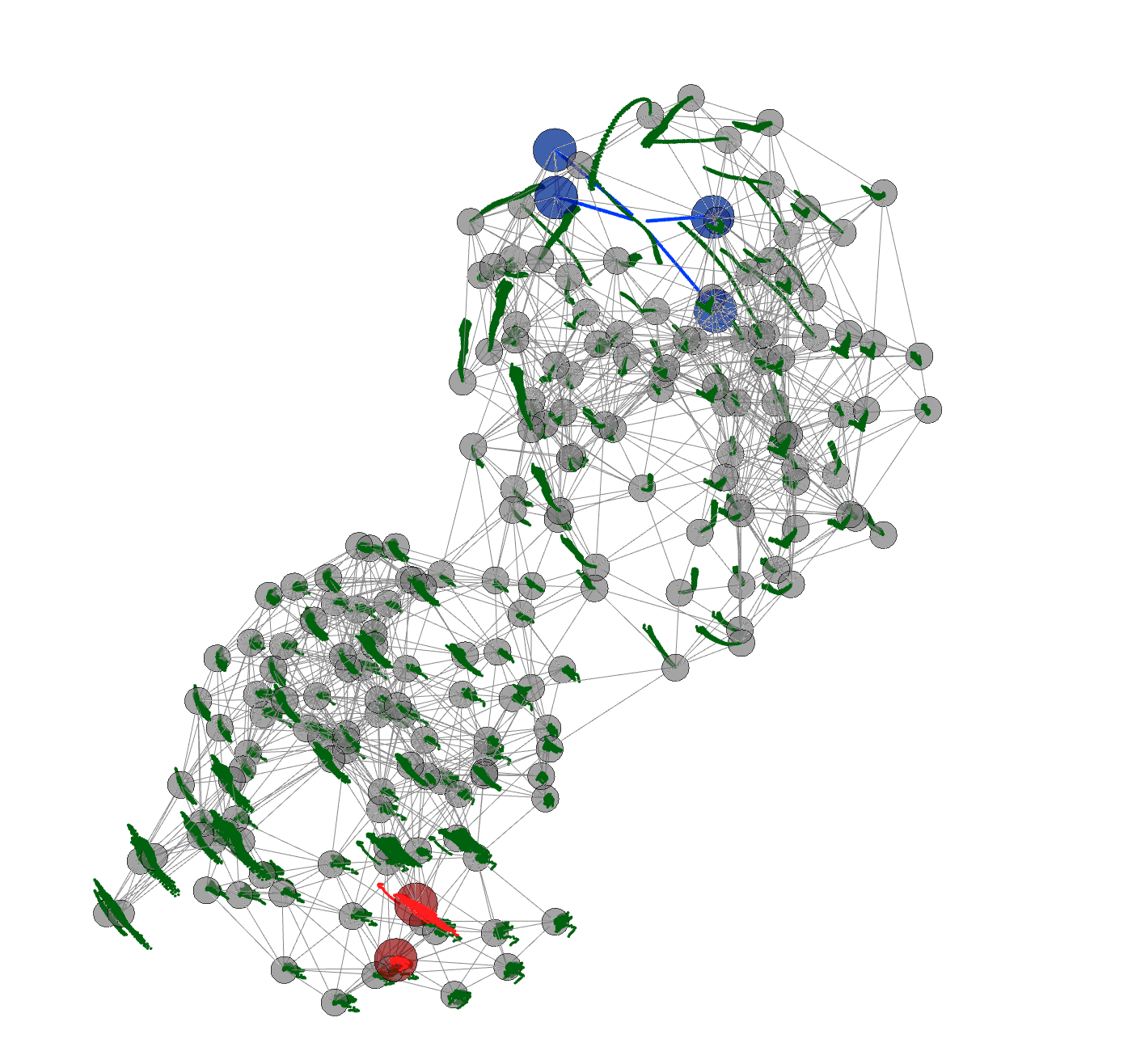}%
  \blue{
    \caption{\textbf{Example of the effect of noise in the input perturbation on the trajectory}
    \label{fig:noise_appendix}
    \textbf{(a-b)} 
    Visualization of multiple response trajectories (green dots) under the influence of normally distributed noise added to the input perturbation at the source site (blue beads). The ensemble of noisy trajectories (green) remains largely confined within a narrow channel around the original response pathway, demonstrating the general robustness of the mechanical response against perturbations. The input noise provides enough energy to occasionally overcome local energy barriers. 
    }
  }
\end{figure*}

\setcounter{equation}{0}
\setcounter{figure}{0}
\setcounter{table}{0}
\renewcommand{\thefigure}{I\arabic{figure}}
\renewcommand{\theequation}{I\arabic{equation}}
\renewcommand{\thetable}{I\arabic{table}}
\blue{
\section{Statistical Analysis of the Conservation Allosteric Sites} \label{Appendix_ks_statistics}
\begin{figure*}
  \includegraphics[width=1.\textwidth]{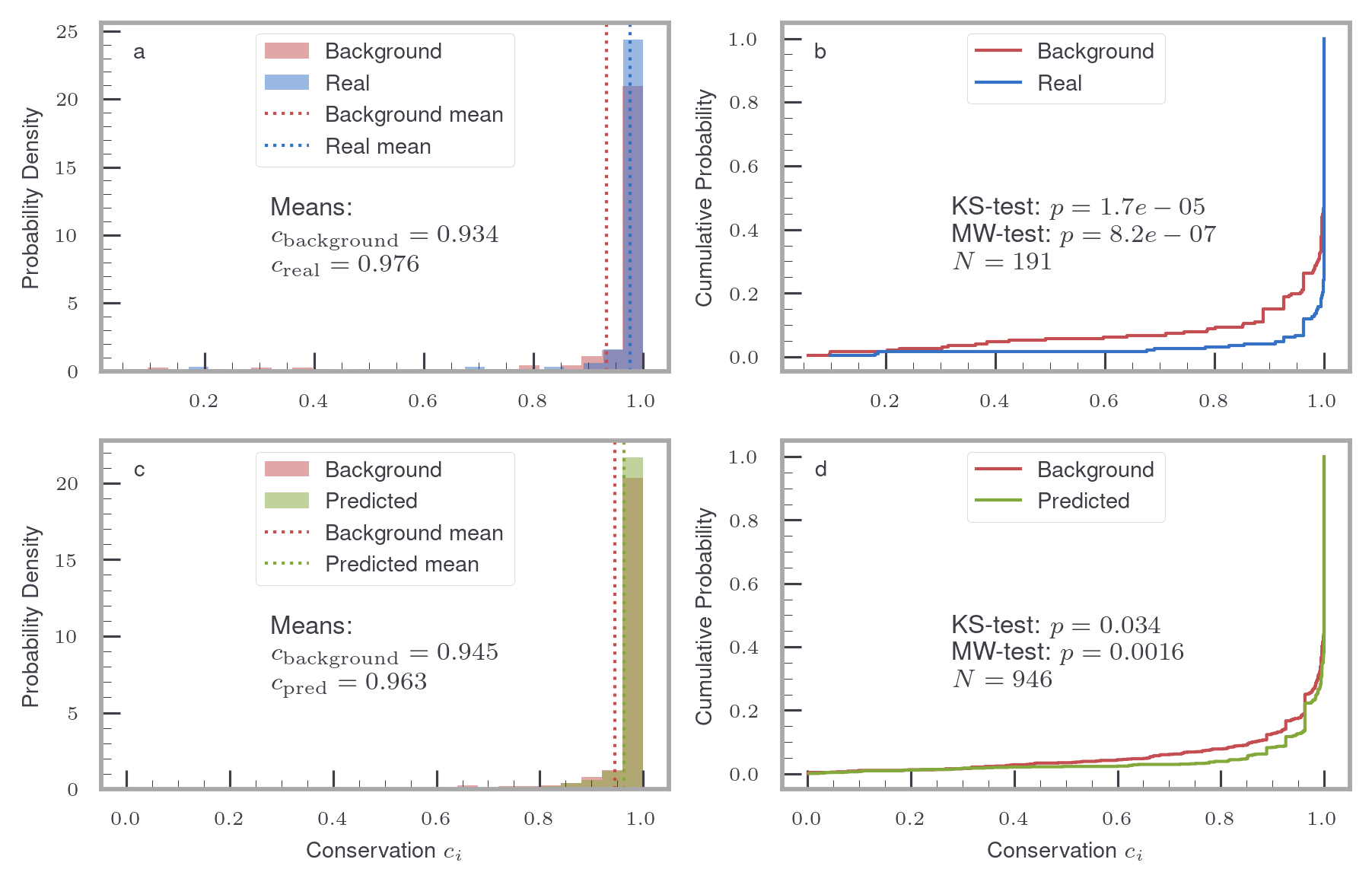}%
    \caption{\blue{\textbf{Statistical analysis of conservation patterns in allosteric sites.}
    Probability density (\textbf{a,c}) and cumulative distribution functions (\textbf{b,d}) of relative conservation values for real allosteric sites (\textbf{a,b}, blue) and predicted allosteric sites (\textbf{c,d}, green) compared to randomly sampled background residues (red). For each comparison, background samples were drawn to match the respective sample sizes (N=191 for real, N=946 for predicted sites). The Kolmogorov-Smirnov test reveals significantly different distributions for both real (p=1.7e-05) and predicted (p=0.034) sites compared to background, indicating distinct conservation patterns. This is supported by the Mann-Whitney U test (p<0.001) which specifically shows that both types of sites are more conserved than background, with real sites showing a stronger conservation signal (mean conservation 0.976) compared to predicted sites (mean conservation 0.963).
    }
  }
  \label{fig:ks_statistics}
\end{figure*}
To statistically validate the conservation patterns, we compared both real and predicted allosteric sites against randomly sampled background residues using two complementary statistical approaches. The Kolmogorov-Smirnov (KS) test was employed to assess whether the overall distributions of conservation values differ significantly. The resulting p-values (p=1.7e-05 for real sites, p=0.034 for predicted sites) below the significance threshold of 0.05 indicate that both real and predicted allosteric sites follow distributions that are significantly different from background conservation patterns. Additionally, we used the Mann-Whitney U test with the alternative hypothesis that allosteric sites show higher conservation than background residues, specifically testing for the directionality of the difference. For each comparison, we drew background samples matching the respective sample sizes to ensure fair statistical comparison. Both the probability density distributions and cumulative distribution functions were analyzed, with the KS statistic representing the maximum distance between the cumulative distribution functions of allosteric and background conservation values.}

\setcounter{equation}{0}
\setcounter{figure}{0}
\setcounter{table}{0}
\renewcommand{\thefigure}{J\arabic{figure}}
\renewcommand{\theequation}{J\arabic{equation}}
\renewcommand{\thetable}{J\arabic{table}}
\blue{
\section{Statistical Analysis of Quantitative Performance Metrics}\label{Appendix_auroc_tables}
To evaluate the performance of our prediction method, we calculated the AUROC values for each protein in our dataset (Table~\ref{tab:auroc}). The AUROC metric ranges from 0 to 1, where 1 indicates perfect prediction and 0.5 represents random chance. Our method achieves consistently high AUROC values across all tested proteins, ranging from 0.720 (2bxd) to 0.946 (4o33), with most values above 0.75. This indicates robust prediction performance across different protein structures. 
}

\begin{table}[h!]
  \centering
  \caption{\blue{Area Under the Receiver Operating Characteristic (AUROC) values for different protein structures.}}
  \label{tab:auroc}  
  \begin{tabular}{lc}
      \hline
      PDB ID & AUROC \\
      \hline \hline
      2bxd & 0.720 \\
      1n5u & 0.799 \\
      1tlf & 0.763 \\
      1wet & 0.783 \\
      4ake & 0.924 \\
      1oe9 & 0.863 \\
      1cmc & 0.721 \\
      4o33 & 0.946 \\
      1ake & 0.852 \\
      2ior & 0.877 \\
      2bxg & 0.794 \\
      1efa & 0.831 \\
      3hrf & 0.765 \\
      1w7j & 0.872 \\
      \hline
  \end{tabular}
\end{table}

  \clearpage
\bibliography{lit}

%
%

\end{document}